\documentclass[aps,prx,onecoloumn,superscriptaddress,12pt]{revtex4-2}
\usepackage[utf8]{inputenc}

\usepackage[english]{babel}
\usepackage{ifthen}
\usepackage{xcolor}
\usepackage{amsmath}
\usepackage{mathtools}
\usepackage{braket}
\usepackage{siunitx}
\usepackage[mode=buildnew]{standalone}
\usepackage{multirow}

\definecolor{darkblue}{rgb}{ 0, 0, 0.7843137254901961}
\definecolor{MOTblue}{rgb}{ 0.7098039215686275, 0.7294117647058823, 1.0 }
\definecolor{darkpurple}{rgb}{0.5568627450980392, 0, 0.7568627450980392}
\definecolor{darkred}{rgb}{0.7568627450980392, 0.15294117647058825, 0.17647058823529413}

\usepackage[bookmarksopen]{hyperref}
\hypersetup{colorlinks,linkcolor=darkblue,citecolor=darkblue,urlcolor=darkblue}

\usepackage{upgreek}

\newcommand{\narrowtimes}{\medmuskip=1mu\times}
\newcommand{\SIangfreq} [2] {\ensuremath{2\pi\narrowtimes\SI{#1}{#2}}}
\DeclareSIUnit[]\bohrradius
{\text{\ensuremath{a_{0} }}}

\DeclareMathOperator\erf{erf}
\DeclareMathOperator\OD{O\!D}

\newcommand{\FigRef}[1]{figure~\ref{#1}}
\newcommand{\FigRefCap}[1]{Figure~\ref{#1}}
\newcommand{\SubFigRef}[2]{figure~\ref{#1}(#2)}
\newcommand{\PSubFigRef}[2]{Figure~\ref{#1}(#2)}
\newcommand{\SubFigRefCap}[2]{Figure~\ref{#1}(#2)}
\newcommand{\EqnRef}[1]{eq.~(\ref{#1})}
\newcommand{\EqnRefCap}[1]{Eq.~(\ref{#1})}
\newcommand{\TblRef}[1]{table~\ref{#1}}
\newcommand{\SecRef}[1]{section~\ref{#1}}

\newcommand{\AffQVIL}[0]{Quantum Valley Ideas Laboratories, 485 Wes Graham Way, Waterloo, ON N2L 0A7, Canada}
\newcommand{\AffPI}[0]{5. Physikalisches Institut, Universität Stuttgart, Pfaffenwaldring 57, 70569 Stuttgart, Germany}

\renewcommand{\Im}{\mathrm{Im}}

\newboolean{ShowComments}\setboolean{ShowComments}{true}  
\provideboolean{ShowComments}

\newcommand{\ShowMyCommnt}[1]{%
\ifShowComments%
#1%
\fi%
}
\newcommand{\NewComment}[2]{%
\expandafter\newcommand\csname#1Comment\endcsname[1]{\ShowMyCommnt{\textcolor{#2}{##1}}}%
}
\NewComment{Harald}{violet}
\NewComment{Jim}{blue}
\NewComment{Matze}{red}
\usepackage[english]{babel}

\begin{document}
\title{Rydberg atom-based radio frequency sensors: amplitude regime sensing}
\author{Matthias Schmidt}
\affiliation{\AffQVIL}
\affiliation{\AffPI}
\author{Stephanie M. Bohaichuk}
\affiliation{\AffQVIL}
\author{Vijin Venu}
\affiliation{\AffQVIL}
\author{Florian Christaller} 
\affiliation{\AffQVIL}
\author{Chang Liu}
\affiliation{\AffQVIL}
\author{Fabian Ripka}
\affiliation{\AffQVIL}
\author{Harald Kübler}
\affiliation{\AffQVIL}
\affiliation{\AffPI}
\author{James P. Shaffer}
\email{jshaffer@qvil.ca}
\affiliation{\AffQVIL}


\begin{abstract}
Rydberg atom-based radio frequency electromagnetic field sensors are drawing wide-spread interest because of their unique properties, such as small size, dielectric construction, and self-calibration. These photonic sensors use lasers to prepare atoms and read out the atomic response to a radio frequency electromagnetic field based on electromagnetically induced transparency, or related phenomena. Much of the theoretical work has focused on the Autler-Townes splitting induced by the radio frequency wave. The amplitude regime, where the change in transmission observed on resonance is measured to determine electric field strength, has received less attention. In this paper, we deliver analytic expressions that are useful for calculating the absorption coefficient and sensitivity in the amplitude regime. We describe the approximations that we applied to obtain the analytic expressions and demonstrate their validity over a large range of the interesting parameter space. The effect of the thermal motion of the atoms is explicitly addressed. Residual Doppler shifts are shown to limit sensitivity.  An analytic expression for the amplitude regime of Rydberg atom-based sensing has not, to our knowledge, been obtained previously. The expressions, approximations and descriptions presented in the paper are important for maximizing the sensitivity of Rydberg atom-based sensors and for providing insight into the physics of multi-level interference phenomena.
\end{abstract}
\maketitle
\section{Introduction}

Atoms provide fundamental advantages in early-stage quantum technologies because all atoms of a particular species have the same properties, whereas uniform fabrication of artificial quantum objects is currently difficult, if not impossible~\cite{Adams_2019, Kuebler2018, Ripka2021}. Atomic Rydberg states can advance quantum technologies towards real-world applications by making use of their sensitivity and tunability ~\cite{Adams_2019}. Rydberg atoms are used in quantum sensing and computation because of their sensitivity to the presence of other atoms and external electromagnetic fields~\cite{Adams_2019, Kuebler2018, Ripka2021}.
The Rydberg manifold consists of many states that are accessible via optical and radio frequency (RF) transitions. The properties of each state scale with principal quantum number, $n$, and angular momentum quantum number, $l$, e.g., the transition dipole moments between neighboring states scale as $n^2$, leading to large resonant enhancements. A broad carrier frequency range spanning MHz-THz is available for RF electric field (E-field) sensing. These characteristics make RF electrometry a promising application of Rydberg atoms~\cite{Sedlacek2012, Holloway2014, Fan_2015, Holloway2017, Simons2021}. Using atoms for RF E-field sensing enables new functionalities like sub-wavelength imaging~\cite{Fan2014,Holloway2017, Ripka2021}, self-calibration and the construction of sensors of practically any shape and size using vapor cells that are electromagnetically transparent~\cite{Fan_2015,Adams2020,Holloway2022}. 

Rydberg atom-based RF E-field sensing relies on measuring spectroscopic changes induced by an incident RF E-field. The transmission changes of a probe laser beam passing through an atomic vapor are used to determine the properties of the RF E-field~\cite{Sedlacek2012}.
Laser light is used to couple the ground state to a Rydberg state in which the atomic electron can (near) resonantly interact with an RF E-field. The atomic-RF E-field interaction modulates the scattering of light (near) resonant with the probe transition, usually the D1 or D2 transition of either rubidium or cesium contained in a room temperature vapor cell. The typical mechanism for coupling the Rydberg state to the probe laser readout is electromagnetically induced transparency (EIT), or related phenomena like electromagnetically induced absorption, because it is sub-Doppler and coherent~\cite{GeaBanacloche1995, Fleischhauer2005}. One helpful interpretation is that the optical fields are used to create a 'new hybrid, 2-level atom', i.e., an optically dressed atom, whose absorption properties are changed in the presence of a RF electromagnetic wave. 
Since the modulated probe laser readout occurs on a strong optical transition, the RF E-field greatly affects the probe laser light transmitted through the vapor cell.

So far, E-field amplitude measurements down to $\SI{\sim 1}{\micro\volt\per\centi\meter}$  with a sensitivity of $\SI{240}{\nano\volt\per\centi\meter\per\hertz\tothe{1/2}}$ have been realised with continuous and pulsed RF E-fields using only optical fields for detection~\cite{Kumar2016, Kumar2017, Kuebler2018, Bohaichuk2022, Sapiro_2020}.  Higher sensitivities have been obtained using an auxiliary RF E-field in a heterodyne measurement~\cite{Bang2022, Dixon2022}.
The progress in pulsed detection has made Rydberg sensors interesting for applications using modulated RF E-fields~\cite{Meyer2018, Anderson2021, Jiao_2019, Holloway2021AMR, Song2019,Bohaichuk2022}. Detection of polarization and phase of the RF electromagnetic wave have been investigated, showing that all the properties of an RF electromagnetic wave can be measured using Rydberg atom-based RF sensors~\cite{Sedlacek2013, Simons2019, Anderson2019RydbergAF, Adams2020}.

To obtain the highest sensitivity and detect the weakest possible RF E-fields without using an auxiliary RF E-field, Rydberg atom sensors can be operated in the amplitude regime. In this mode, the RF E-field is determined by a relative change of the optical absorption on resonance~\cite{Sedlacek2012,Kuebler2018, Ripka2021}. Despite the utility of the amplitude regime, it has attracted little theoretical attention. In this paper, we present a comprehensive theoretical study that includes simplified expressions for the absorption coefficient in the presence of the RF E-field that are valid over a wide parameter range, including an expression that accounts for Doppler shifts. 
We compare the cases of stationary atoms, perfect wave vector matching of the excitation lasers and room temperature cesium atoms with a finite wave vector mismatch. Several straightforward formulas are presented that show how the laser and atomic decay parameters are related to the RF E-field induced change in absorption. The results enable the recalculation of the incident RF E-field strength given a measured absorption change. Although not as elegant as the Autler-Townes regime, the RF E-field strength can be determined provided the coupling laser Rabi frequency is known. Our work is important for choosing experimental parameters that maximize the sensitivity of Rydberg atom-based sensors, particularly in cases where it is advantageous to avoid using an auxiliary RF E-field, and for providing insight into the physics of multi-level interference phenomena.

\section{Methods}\label{sec:Introduction}

\begin{figure}
\centering
\includegraphics[width=\columnwidth]{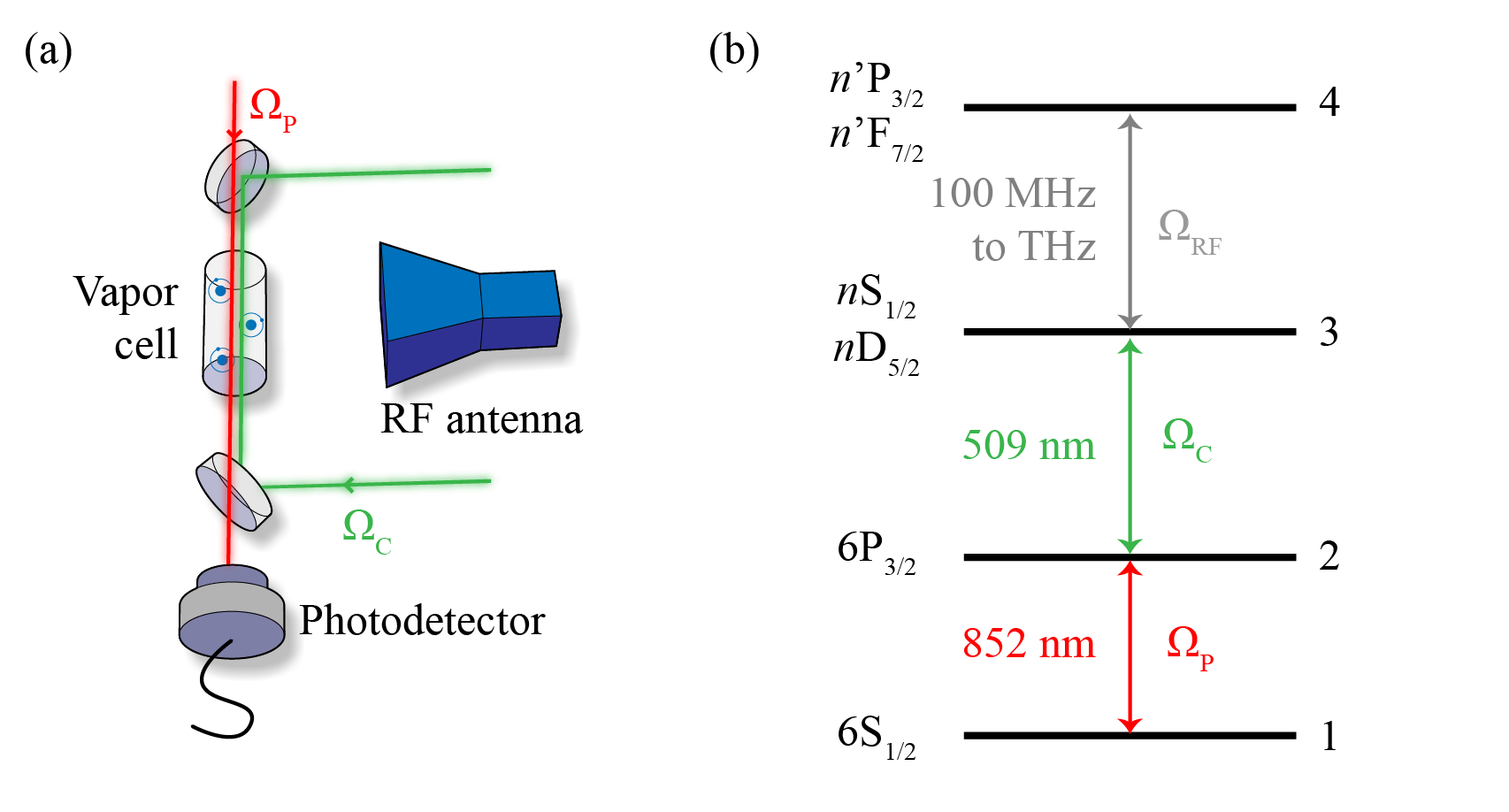}
\caption{(a) Schematic of a simplified experimental setup for Rydberg atom-based electrometry. Two counter-propagating laser beams, the probe and the coupling laser, denoted by their respective Rabi frequencies, $\Omega_\mathrm{P}$ and $\Omega_\mathrm{C}$, pass through a dielectric glass vapor cell filled with cesium atoms at room temperature. The RF electromagnetic wave is emitted by a RF source, e.g., an antenna. The transmission of the probe beam is detected on the photodetector. (b) Excitation scheme used for detection of the RF E-field. The probe laser drives the D2 transition of cesium with Rabi frequency $\Omega_\mathrm{P}$. A coupling laser drives the transition from the intermediate to the launch Rydberg state, i.e. the Rydberg state participating in EIT in the absence of the RF E-field, with Rabi frequency $\Omega_\mathrm{C}$. The incident RF wave, $\Omega_\mathrm{RF}$, couples two Rydberg states and alters the optical transmission of the probe laser passing through the vapor cell.}
\label{Fig:ExperimentalSetupLevelScheme}
\end{figure}

A representative setup for RF E-field sensing using atoms in Rydberg states consists of a dielectric vapor cell filled with cesium or rubidium vapor at room temperature, several lasers and a photodetector. \PSubFigRef{Fig:ExperimentalSetupLevelScheme}{a} shows an example where two laser beams, a probe and a coupling beam, pass through the vapor cell in a counter-propagating geometry. The change in transmission of the probe laser beam is read out optically by a photodetector after passing through the vapor cell. The local RF E-field in combination with the coupling laser alters the atomic response to the probe laser field. We consider the case of Rydberg atom electrometry with a cesium vapor. Experiments where there are three lasers have been explored \cite{Kuebler2018, Ripka2021, Bohaichuk2022, Ripka2022} and cases where the two laser beams co-propagate are possible. The theory presented in this work can be adapted to these other configurations in a straightforward manner.

The optical response of the atom can be calculated by solving the Lindblad master equation using the density matrix formalism. Since we are primarily concerned with the scattering of light out of the probe laser beam, the $\rho_{12}$ matrix element is of central focus. $\rho_{12}$ can be used to calculate the expectation value of the induced transition dipole moment with which the probe laser interacts with the atoms, 
\begin{equation}
\langle\hat{\mu}_\mathrm{P}\rangle = \mu_\mathrm{P} \rho_{21}+\mu_\mathrm{P}^\star\rho_{12},
\end{equation}
where $\hat{\mu}_\mathrm{P}$ is the dipole operator for the probe laser transition and $\mu_{P} = \bra{1}\hat{\mu}_\mathrm{P}\ket{2}$. To determine $\rho_{12}$, the Lindblad master equation, 
\begin{equation}
\dot{\rho}=\frac{i}{\hbar} [\cal{H}, \rho]-\cal{L},
\label{Eq:Theory_DensityMatrixEquation}
\end{equation}
is solved in the steady-state limit, where $\hbar$ is the reduced Planck constant.
The Hamiltonian, $\cal{H}$, describes the coupling of the laser light fields and the RF E-field to the atoms. The choice of near resonant laser transitions typically leads to a ladder excitation scheme like the one we use for our analysis, shown in \SubFigRef{Fig:ExperimentalSetupLevelScheme}{b}. $\cal{H}$ can be expressed for the ladder scheme as a matrix
\begin{equation}
{\cal{H}}=\frac{\hbar}{2}\left(\begin{array}{cccc}
0 & \Omega_\mathrm{P} & 0 & 0  \\
\Omega_\mathrm{P} & -2 \Delta_\mathrm{P} & \Omega_\mathrm{C} & 0  \\
0 & \Omega_\mathrm{C} & -2\left(\Delta_\mathrm{P}+\Delta_\mathrm{C}\right) & \Omega_\mathrm{RF}  \\
0 & 0 & \Omega_\mathrm{RF} & -2\left(\Delta_\mathrm{P}+\Delta_\mathrm{C}+\Delta_\mathrm{RF}\right)
\end{array}\right).
\label{Eq:Theory_Hamiltonian}
\end{equation}
The Rabi frequencies of the individual driving fields are given by $\Omega_i = \mu_{i}E_i/\hbar$, where $i$ labels the probe (P), coupling (C) or radio frequency (RF) transition. The dipole matrix element for each transition is denoted $\mu_{i}$. $E_i$ denotes the amplitude of each field. $\Delta_i$ is the detuning of the respective field from its resonance. The RF E-field is assumed to be on resonance, $\Delta_\mathrm{RF} = 0$, throughout the paper. The probe laser transition is chosen to drive the D2 transition of cesium at $\lambda_\mathrm{P} = \SI{852}{\nano \meter}$ while the coupling laser wavelength is chosen to be at $\lambda_\mathrm{C} = \SI{509}{\nano \meter}$. 
The Lindblad operator, $\cal{L}$, describes the decay and dephasing of the populations and coherences~\cite{Kuebler2018, Bohaichuk2022}. In our numerical calculations, we use $\gamma_{2} = \SIangfreq{5}{\mega \hertz}$ for the decay rate of the $6P_{3/2}$ state. For both Rydberg states, we assume total decay rates of $\gamma_3 = \gamma_4 = \SIangfreq{20}{\kilo \hertz}$. The values for $\gamma_3$ and $\gamma_4$ are used as representative decay rates for a Rydberg state. Collisional dephasing and laser dephasing are neglected explicitly in our calculations to simplify the picture, but can be introduced in the Lindblad operator in a straightforward manner. Transit time broadening is also neglected for most of the paper. We include transit time broadening in the sensitivity calculations to get realistic results.

In a finite temperature vapor, the atoms move with a distribution of different velocities. 
The atoms experience Doppler shifts that are proportional to their velocity, $v$, that cannot be neglected in most Rydberg atom vapor cell electrometers. The Doppler shifts appear in the density matrix equations as velocity dependent detunings,
\begin{equation}
\Delta_i =\vec{k}_i \cdot \vec{v},
\label{Eq:Thermal_DopplerShift}
\end{equation}
where $|\vec{k}_i|= 2\pi/\lambda_i$ denotes the magnitude of the wave vector of the laser whose wavelength is $\lambda_i$. Since the lasers are co-linear in our geometry, \EqnRef{Eq:Thermal_DopplerShift} is a scalar equation where the sign of $k_i$ denotes the propagation direction of the lasers relative to one another and only the atomic velocity along the laser beam directions needs to be taken into account. The Doppler shifts associated with the RF transition are small compared to those of the probe and coupling laser, because $\lambda_\mathrm{P}, \lambda_\mathrm{C} \ll \lambda_\mathrm{RF}$. 
We include the Doppler shifts, as described in \EqnRef{Eq:Thermal_DopplerShift}, for the probe and coupling laser transitions in the Hamiltonian, \EqnRef{Eq:Theory_Hamiltonian}, but neglect those from the RF transition. The density matrix equations yield matrix elements that generally depend on the atomic velocity. 

Later, we assume the weak probe approximation from which we conclude that the absorption coefficient, $\alpha$, is independent of the probe laser intensity. The transmission of the probe laser beam through the vapor cell can then be described by the Lambert-Beer law
\begin{equation}
I(z) = I_0 \mathrm{e}^{-\alpha z},
\label{BeerLaw}
\end{equation}
where the initial probe laser intensity, $I_0$, exponentially decreases with the propagation distance, $z$, through the medium. The Lambert-Beer law also describes the absorption of the other fields in the vapor cell, but it is the probe laser that is typically read out in Rydberg atom electrometry. 
In the weak probe approximation, the ground state is predominantly populated, so absorption of the coupling laser and RF E-field are insignificant in comparison to the probe laser.

The absorption coefficient, $\alpha$, describes the optical response of the atoms observed in experiments. $\alpha$ depends on the probe laser field, the coupling laser field and the RF E-field, including their characteristics like amplitude and detuning. $\alpha$ is directly proportional to the transition dipole moment of the probe transition. The absorption coefficient is defined as 
\begin{equation}
\alpha = \frac{3\pi  n \lambdabar^3_\mathrm{P} }{2  }  k_\mathrm{P} \frac{\gamma_2}{\Omega_\mathrm{P}}  \Im(\rho_{12}) = A k_\mathrm{P} \frac{\gamma_2}{\Omega_\mathrm{P}} \Im (\rho_{12}).
\label{AbsorptionCoefficientDefinition}
\end{equation}
The density of atoms, $n$, can be calculated using a temperature dependent vapor pressure model~\cite{Steck2019}. The reduced wavelength of a photon emitted on the probe transition is $\lambdabar_\mathrm{P} = \lambda_\mathrm{P}/2\pi $. The scattering of photons on the probe transition is modified by the incident RF electromagnetic waves interacting with the atoms through $\rho_\mathrm{12}$. $A$ is proportional to the number of atoms in a probe transition wavelength mode volume. The fact that the single atom density matrix can describe the atomic response indicates that each atom is an independent sensor.

The absorption coefficient measured in experiments is calculated by integrating over all atomic velocity classes weighted by the Maxwell-Boltzmann distribution for the thermal gas. The absorption coefficient of the thermal ensemble of atoms is
\begin{equation}
\alpha = A k_\mathrm{P} \frac{\gamma_2}{\Omega_\mathrm{P}} \int_{-\infty}^{\infty} P(v) \Im(\rho_{12}(v)) \, \mathrm{d}v,
\label{Eq:GeneralAlphaCalcThermal}
\end{equation}
where the Maxwell-Boltzmann velocity distribution of the gas is
\begin{equation}
P(v)=\sqrt{\frac{m}{2\pi k_\mathrm{B}T}} \exp\left(\frac{-mv^2}{2k_\mathrm{B}T}\right) =\frac{1}{ \sqrt{\pi}\bar{v}} \exp\left(\frac{- v^2}{\bar{v}^2}\right).
\label{Eq:Therm-Dist}
\end{equation}
$m$ is the atomic mass, $k_\mathrm{B}$ is the Boltzmann constant, and $T$ is the temperature of the gas. The average velocity of the thermal cloud is $\bar{v}=\sqrt{2k_\mathrm{B} T/m}$. We use room temperature, $T=\SI{20}{\celsius}$, for the calculations, unless otherwise noted. Determining straightforward and instructive expressions for $\alpha$ on resonance under realistic conditions is the main goal of the work.

\begin{figure*}
\includestandalone[width = \textwidth]{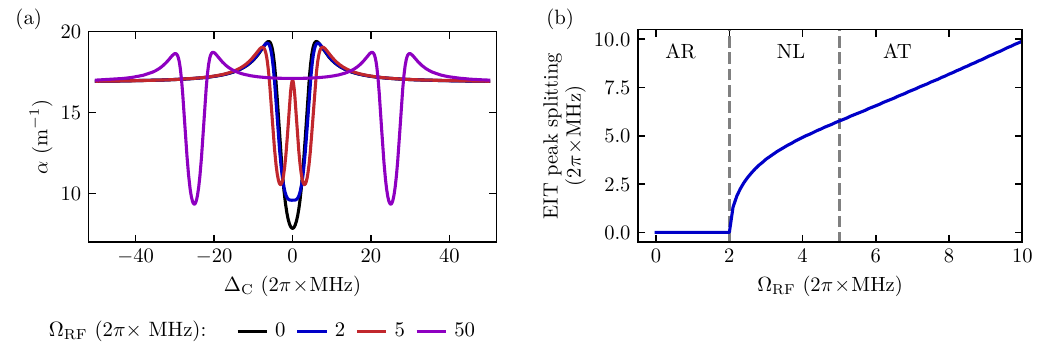}
\caption{(a) Probe laser absorption coefficient, $\alpha$, at $\Omega_\mathrm{P}=\SIangfreq{100}{\kilo \hertz}$ as a function of $\Delta_\mathrm{C}$, for $\Omega_\mathrm{C}= \SIangfreq{5}{\mega \hertz}$. The spectra are displayed for a selection of different $\Omega_\mathrm{RF}$.
(b) At $\Omega_\mathrm{RF}$ between $\SIangfreq{2}{\mega \hertz}$ and $\SIangfreq{5}{\mega \hertz}$ the transmission window splitting behaves non-linearly (NL). At $\Omega_\mathrm{RF} < \SIangfreq{2}{\mega \hertz}$, the presence of the RF E-field is detectable by a change in probe laser transmission on resonance (AR). We define the AT boundary to be where the RF E-field induced spectral splitting is larger than the EIT linewidth,
$\Omega_\mathrm{RF}>\SIangfreq{5}{\mega \hertz}$.
}
\label{Fig:SpectrumEITSplitting}
\end{figure*}

\section{Amplitude Regime}

A typical set of calculated spectra for RF Rydberg atom-based sensors is shown in \SubFigRef{Fig:SpectrumEITSplitting}{a}. The probe laser absorption coefficient, $\alpha$, is shown as a function of the coupling laser detuning, $\Delta_\mathrm{C}$, for different $\Omega_\mathrm{RF}$. The probe laser Rabi frequency is $\Omega_\mathrm{P}= \SIangfreq{100}{\kilo \hertz}$, while the coupling laser Rabi frequency is $\Omega_\mathrm{C} = \SIangfreq{5}{\mega \hertz}$. The probe laser is resonant, $\Delta_\mathrm{P} = 0$. For $\Omega_\mathrm{RF}=0$, the spectrum shows a distinct transmission feature as the coupling laser is scanned across the atomic resonance due to EIT. The spectral width of the transmission feature is determined by Doppler and power broadening. 
In the presence of weak $\Omega_\mathrm{RF} < \SIangfreq{2}{\mega \hertz}$, the medium becomes more absorptive around the EIT resonance~\cite{Sedlacek2012,Kuebler2018}. For larger $\Omega_\mathrm{RF}\approx \SIangfreq{3}{\mega \hertz}$, the EIT feature splits, indicating Autler-Townes (AT) splitting. 

The effect of the RF E-field on the resonant probe laser transmission is illustrated in \SubFigRef{Fig:SpectrumEITSplitting}{b}. Three regions can be identified as the magnitude of $\Omega_\mathrm{RF}$ changes. At large $\Omega_\mathrm{RF}$, the splitting is proportional to $\Omega_\mathrm{RF}$. This is the 'so-called' Autler-Townes regime (AT). The AT is reached when $\Omega_\mathrm{RF}$ is larger than the spectral line width of the EIT feature. In \SubFigRef{Fig:SpectrumEITSplitting}{b}, the AT begins around $\Omega_\mathrm{RF}\approx \SIangfreq{5}{\mega \hertz}$. The AT has been the subject of several studies~\cite{Fleischhauer2005, Sedlacek2012, Gordon2014, Hao2018, Adams2020}. When $\Omega_\mathrm{RF}$ is smaller than the linewidth of the EIT transmission feature, the spectral splitting is not linear in $\Omega_\mathrm{RF}$. We refer to this as the nonlinear regime (NL). For the smallest $\Omega_\mathrm{RF}$, the splitting is not resolvable. The strength of the RF E-field in this limit is detected as a change in the EIT transmission on resonance. The region where the RF E-field induced splitting is much less than the EIT line width is referred to as the amplitude regime (AR) and is the subject of the following sections of the paper.

\section{The Weak Probe Approximation}\label{Sec:WeakProbe}

To simplify the expressions obtained from \EqnRef{Eq:Theory_DensityMatrixEquation}, we adopt the weak probe approximation. In the weak probe limit, $\Omega_\mathrm{P}\ll \gamma_2,\Omega_\mathrm{C}$, the dependence of $\rho_{12}$ on $\Omega_\mathrm{P}$ is linear as terms of higher order in $\Omega_\mathrm{P}$ are neglected~\cite{Firstenberg2016}. In this approximation, the atomic population is predominantly in the ground state and $\rho_{12}$ can  be expressed as the product of two terms, $L^\prime$ and $S$,
\begin{equation}
    \rho_{12} = L^\prime S= \frac{\Omega_\mathrm{P}}{\gamma_2} L S = C_\mathrm{P} L S. 
    \label{Eq:OpticalResponse4Level}
\end{equation}
The first term, $L^\prime$, describes the absorption of the probe laser beam in the absence of both the coupling laser and RF E-field. $L$ has the form of a two-level system,
\begin{equation}
L= \frac{i \gamma_2}{\gamma_2 + 2i \Delta_\mathrm{P}}=
\frac{1}{2\delta_\mathrm{P}-i}=
\frac{2 \delta_\mathrm{P}}{1+4 \delta_\mathrm{P}^2}+ \frac{i}{1+4\delta_\mathrm{P}^2}.
\label{Eq:L}
\end{equation}
In \EqnRef{Eq:OpticalResponse4Level} and \EqnRef{Eq:L}, we introduce the coherence parameter for the probe laser transition, $C_\mathrm{P}= \Omega_\mathrm{P}/\gamma_2$, and a normalized probe laser detuning, $\delta_\mathrm{P} = \Delta_\mathrm{P}/\gamma_2$. The coherence parameter compares the coherent driving of the atomic response to the dephasing. The coherence parameter is a measure of the number of Rabi oscillations that occur during the decay time of the excited state. The imaginary part of $L$ yields a Lorentzian absorption curve as a function of $\Delta_\mathrm{P}$. The Lorentzian width is proportional to $\gamma_2$. In the weak probe limit, $\alpha$ becomes independent of the probe beam power, see \EqnRef{AbsorptionCoefficientDefinition}.

The coupling laser and RF E-fields modify the probe laser absorption through $S$,
\begin{equation}
S =\dfrac{1}{1+\dfrac{\dfrac{\Omega^2_\mathrm{C}}{(\gamma_2+2i\Delta_\mathrm{P})(\gamma_3 + 2i( \Delta_\mathrm{P}+\Delta_\mathrm{C}))}}{1+ \dfrac{\Omega^2_\mathrm{RF}}{(\gamma_3 + 2i(\Delta_\mathrm{P}+\Delta_\mathrm{C}))(\gamma_4 + 2i(\Delta_\mathrm{P}+ \Delta_\mathrm{C} + \Delta_\mathrm{RF}))}}}.
\label{Eq:Slong}
\end{equation}
To simplify the form of \EqnRef{Eq:Slong}, we introduce the normalized two-photon detuning, $\delta_\mathrm{C} = (\Delta_\mathrm{P}+\Delta_\mathrm{C} )/\gamma_3$, the normalized three-photon detuning, $\delta_\mathrm{RF} = (\Delta_\mathrm{P}+\Delta_\mathrm{C}+\Delta_\mathrm{RF})/\gamma_4$, the coupling laser transition coherence, $C_\mathrm{C}= \Omega_\mathrm{C}/\sqrt{\gamma_2\gamma_3}$, and the RF transition coherence, $C_\mathrm{RF}= \Omega_\mathrm{RF}/\sqrt{\gamma_3\gamma_4}$. Our definitions assume phase insensitive transitions, $\Omega = \Omega^\star$. These quantities mirror $C_\mathrm{P}$, introduced in $L$. A more compact form for $S$ obtained by subsitution of the coherences and normalized detunings is
\begin{equation}
S =\dfrac{1}{1-\dfrac{\dfrac{C^2_\mathrm{C}}{(2\delta_\mathrm{P}-i)( 2\delta_\mathrm{C}-i)}}{1- \dfrac{C^2_\mathrm{RF}}{( 2\delta_\mathrm{C}-i)( 2\delta_\mathrm{RF}-i)}}}.
\label{Eq:S}
\end{equation}
The resonance features due to the coupling laser and RF field can be clearly identified in \EqnRef{Eq:S}. The respective coherences appear separated in each of the fractions appearing in the denominator. Each coherence is compared to its respective field detuning. The effect of the RF E-field on the probe laser absorption is transferred via the coupling laser coherence to the probe laser transition. If we first consider $\Omega_\mathrm{RF} =0$ and all fields resonant, EIT transmission appears due to $\Omega_\mathrm{C}$. As $C_\mathrm{C}$ becomes much larger than one, the probe laser absorption is directly modulated by the coupling laser Rabi frequency, $\Omega_\mathrm{C}$, decreasing as $\Omega_\mathrm{C}$ increases. If a finite $C_\mathrm{RF}$ is introduced, it can reduce the response of the atom due to $C_\mathrm{C}$, increasing the probe absorption. Generally, for a Rydberg transition, $C_\mathrm{RF}$ has a high degree of coherence since $\Omega_\mathrm{RF} \gg \gamma_3,\gamma_4$. However, at the lowest detectable RF E-fields, the inequality can be violated, reducing the influence of $\Omega_\mathrm{RF}$ on $S$. Likewise, a highly incoherent RF electromagnetic wave, $C_\mathrm{RF}\ll 1$, will have little effect on the probe light scattering.

The picture we described can be extended to any ladder excitation scheme. In the weak probe approximation, the absorption coefficient can be expressed as the product of a probe laser transition Lorentzian $L$ and a modulating signal $S$ that changes the two-level optical response of the atom interacting with the probe laser. 
The coupling laser can suppress the absorption of the probe laser, consistent with EIT. The RF E-field acts on the coupling laser field to reduce its effect on the probe laser scattering, increasing absorption. Processes like collisions and blackbody decay can be accounted for by adding modifications to $\cal{L}$ in most cases of interest. Since Rydberg atom dephasing generally impacts Rydberg atom-based sensors negatively, these processes are minimized in experiments.

\section{The Doppler Effect}\label{Sec:Doppler effect}

In \SecRef{Sec:WeakProbe}, we did not explicitly address how atomic motion modifies the atomic response. Since one of the great advantages of Rydberg atom-based sensors is their room temperature operation, the Doppler effect cannot be ignored. For RF E-field sensing applications in the AR, a large signal-to-noise ratio as well as a large slope with respect to changes in $\Omega_\mathrm{RF}$ are advantageous. The amplitude of the EIT transmission feature and the change in $\alpha$ with respect to $\Omega_\mathrm{RF}$ depend on the magnitude and sign of $k_\mathrm{eff} = k_\mathrm{P} + k_\mathrm{C}$ and $k_\mathrm{P}$. 
Neglecting $k_\mathrm{RF}$ is an approximation, but the RF wavelength is long compared to wavelengths of the laser fields. The Doppler effect plays an important role in determining absorption amplitude and the spectral line width. 

\begin{figure*}
\centering
\includestandalone[width = \textwidth]{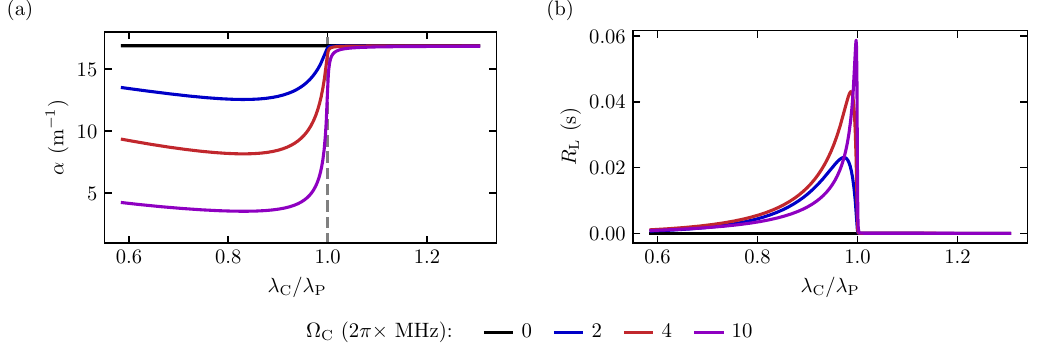}
\caption{(a) Absorption coefficient as a function of the relative wavelength $\lambda_\mathrm{C}/\lambda_\mathrm{P}$ for various $\Omega_\mathrm{C}$ at $\Omega_\mathrm{RF}=0$. For $\lambda_\mathrm{C}>\lambda_\mathrm{P}$, the EIT signal almost vanishes completely. (b) Logarithmic derivative, $R_\mathrm{L}$, as a function of $\lambda_\mathrm{C}/\lambda_\mathrm{P}$ for $\Omega_\mathrm{RF} = \SIangfreq{0.5}{\mega \hertz}$. For $\lambda_\mathrm{C} > \lambda_\mathrm{P}$, the atoms are insensitive to a change in $\Omega_\mathrm{RF}$. $\Omega_\mathrm{P} = \SIangfreq{100}{\kilo \hertz}$.}
\label{Fig:DeltaK_Zero_SignalHeight}
\end{figure*}

\SubFigRefCap{Fig:DeltaK_Zero_SignalHeight}{a} shows the absorption amplitude as a function of the relative wavelength $\lambda_\mathrm{C}/\lambda_\mathrm{P}$ for different $\Omega_\mathrm{C}$ at $\Omega_\mathrm{RF}=0$. The probe laser absorption, i.e., $\Omega_\mathrm{C}=0$, is shown as a black line. The curves all show similar behavior. For $\lambda_\mathrm{C}> \lambda_\mathrm{P}$, the EIT signal is weak. For $\lambda_\mathrm{C} < \lambda_\mathrm{P}$, absorption minima exist as a function of $\lambda_\mathrm{C}/\lambda_\mathrm{P}$. There is an optimum wavelength mismatch to obtain maximum EIT amplitude. The optimum wavelength mismatch is determined by the summation of the contributions to the absorption from each atom in the Doppler distribution. When $\lambda_\mathrm{C}/\lambda_\mathrm{P}$ exceeds one, the relative sign of the two-photon Doppler shift and the probe laser Doppler shift changes. In the region where $\lambda_\mathrm{C}/\lambda_\mathrm{P} > 1$ the Doppler shifts add, rather than subtract from each other, causing suppression of the EIT signal~\cite{Urvoy2013}. The strong increase in absorption is caused by the fact that the detuning of the AT-split lines from the bare atomic states cannot be compensated by the two photon Doppler shift.

For the excitation scheme presented in \SubFigRef{Fig:ExperimentalSetupLevelScheme}{b}, with a relative wavelength $\lambda_\mathrm{C}/\lambda_\mathrm{P} = \SI{509}{\nano \meter} / \SI{852}{\nano \meter} \approx 0.6$, $\alpha(\Omega_\mathrm{RF}=0)$ clearly shows a larger EIT signal than an excitation scheme with $k_\mathrm{eff} =0$ for all $\Omega_\mathrm{C}$. However, as $\lambda_\mathrm{C}/\lambda_\mathrm{P}$ decreases from one, the spectral line width increases due to the residual Doppler shifts arising from a larger effective wave vector mismatch, $k_\mathrm{eff}$. The narrowest spectral line width occurs at $k_\mathrm{eff}=0$. The optimal condition for high sensitivity is found in between these two extremes, at the ideal trade-off between the signal height and the spectral line width.

We define the responsitivity, $R$, as the derivative of $\alpha$ with respect to $\Omega_\mathrm{RF}$. We define the relative absorption coefficient $\Delta \alpha$,
\begin{equation}
\Delta \alpha = \dfrac{\alpha (\Omega_\mathrm{RF})-\alpha(\Omega_\mathrm{RF}=0)}{\alpha(\Omega_\mathrm{RF}=0)}.
\label{Eq:NonThermal_RelAbsorptionCoeff}
\end{equation}
Combining the definitions of $R$ and $\Delta \alpha$,  we can approximate the logarithmic derivative of $\alpha$ with respect to $\Omega_\mathrm{RF}$, 
\begin{equation}
R_\mathrm{L} = \frac{\partial \alpha}{\alpha(\Omega_\mathrm{RF}=0)\, \partial \Omega_\mathrm{RF}} \approx \lim_{\Delta \Omega_\mathrm{RF}\to 0} \frac{\Delta \alpha }{\Delta \Omega_\mathrm{RF}}.
\end{equation}
$\Delta \Omega_\mathrm{RF}$ is the change in the RF E-field, assumed to be small. $R_\mathrm{L}$ is useful because it provides a local measure of the sensitivity that is independent of the coefficient $A$. \SubFigRefCap{Fig:DeltaK_Zero_SignalHeight}{b} shows an approximation to $R_\mathrm{L}$, i.e.,  
\begin{equation}
R_\mathrm{L} = \frac{\alpha(\Omega_\mathrm{RF})-\alpha(\Omega_\mathrm{RF}=0)}{\alpha(\Omega_\mathrm{RF}=0)\, \Omega_\mathrm{RF}},
\end{equation}
where $\Omega_\mathrm{RF} = \SIangfreq{0.5}{\mega \hertz}$.  For $\lambda_\mathrm{C}> \lambda_\mathrm{P}$, the absorption change of the atoms in response to the RF E-field is small. Likewise, as $\lambda_\mathrm{P} \gg \lambda_\mathrm{C}$ the signal decreases as the residual Doppler broadening increases. As $\lambda_\mathrm{C}/ \lambda_\mathrm{P}$ approaches one from the lower limit in \SubFigRef{Fig:DeltaK_Zero_SignalHeight}{b} a maximum is evident. The maximum sensitivity point is a balance between narrow spectral linewidth and signal amplitude. The ideal sensor has nearly matched wavelengths where the coupling laser wavelength is slightly shorter than the probe laser wavelength. \SubFigRefCap{Fig:DeltaK_Zero_SignalHeight}{b} illustrates the importance of the Doppler effect in Rydberg atom-based sensors and points to pursuing the three photon sensing scheme originally proposed by several of the authors~\cite{Kuebler2018} and demonstrated experimentally in~\cite{Bohaichuk2023threephoton}.

To further elucidate how the Doppler effect changes the optical absorption, we consider three cases. First, we address the zero temperature limit, $T=0$, where all atoms are at rest and the Doppler effect is eliminated. Next, we examine the case where there is perfect wave vector matching between the probe and coupling laser fields in a counter-propagating geometry, $k_\mathrm{eff}=0$, and the Doppler effect only influences the probe transition. The first two cases are illustrative and are intended to provide insight into the more complicated case of finite wave vector mismatch and temperature, the realistic experimental case, that is considered last.

\subsection{The Zero Temperature Limit}\label{Sec:Teq0}

To simplify \EqnRef{Eq:OpticalResponse4Level}, we set the probe laser detuning to resonance, $\Delta_\mathrm{P}=0$, and eliminate Doppler shifts by setting $T=0$. In this limit, the probe absorption spectrum obtained by scanning the coupling laser looks similar to \SubFigRef{Fig:SpectrumEITSplitting}{a}. However, the spectral width of the peaks is narrow, limited by decay processes. 

\begin{figure*}
\includestandalone[width = \textwidth]{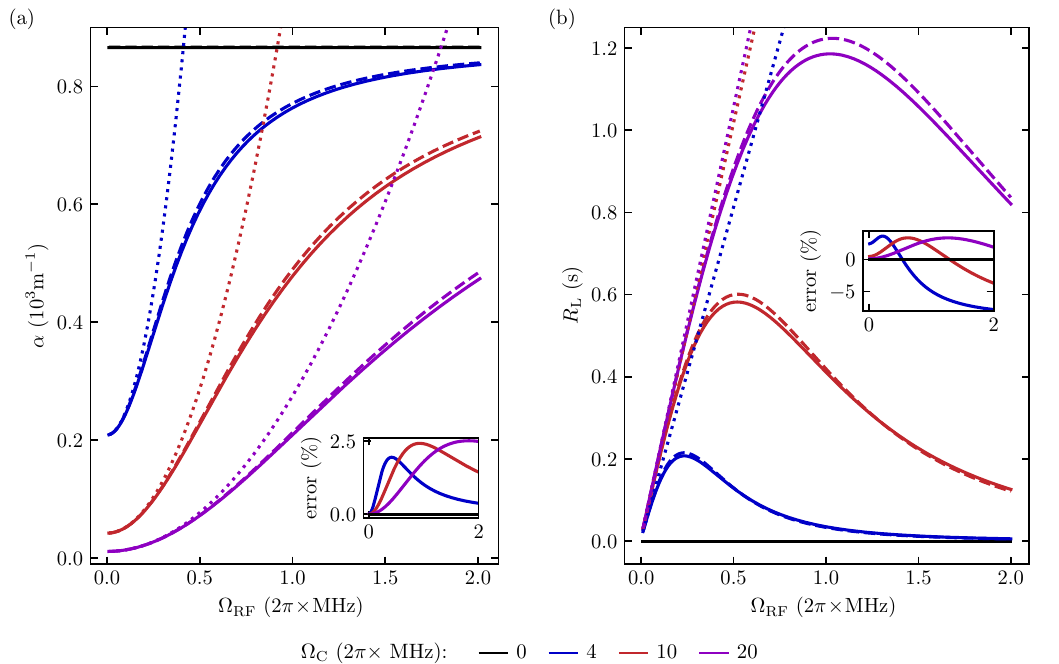}
\caption{(a) On resonant absorption coefficient, $\alpha$, as a function of $\Omega_\mathrm{RF}$ for several $\Omega_\mathrm{C}$ in the AR for $T=0$. $\Omega_\mathrm{P}= \SIangfreq{100}{\kilo \hertz}$. Solid lines represent the full density matrix calculation, dashed lines correspond to the weak probe approximation, \EqnRef{Eq:NonThermal_WeakProbe}, and dotted lines correspond to the strong coupling approximation, \EqnRef{Eq:NonThermal_StrongCoupling}. The relative error between the full density matrix calculation and the weak probe approximation as a function of $\Omega_{RF}$ is shown in the inset.
(b) $R_\mathrm{L}$ as a function of $\Omega_\mathrm{RF}$ for the same $\Omega_\mathrm{C}$ as plotted in (a). The relative error between the full density matrix calculation and the weak probe approximation is shown in the inset of the plots.
}
\label{Fig:NonThermalAbsorptionCoefficientRelRec}
\end{figure*}

In \SubFigRef{Fig:NonThermalAbsorptionCoefficientRelRec}{a} the full density matrix equations are used to calculate the resonant absorption coefficient, $\alpha$ as a function of $\Omega_\mathrm{RF}$ for several $\Omega_\mathrm{C}$. The probe Rabi frequency is ${\Omega_\mathrm{P} = \SIangfreq{100}{\kilo \hertz}}$. The plots exhibit several basic features that are consistent with experimental observations \cite{Sedlacek2012,Fan_2015}. For small $\Omega_\mathrm{RF}$, $\alpha$ increases quadratically with $\Omega_\mathrm{RF}$. At intermediate $\Omega_\mathrm{RF}\geq \SIangfreq{0.5}{\mega \hertz}$, the EIT transmission feature begins to split and the system enters the NL. For larger $\Omega_\mathrm{RF}$, $\alpha$ approaches the background absorption indicating the AT. In the absence of the coupling laser, the Rydberg state is empty and the vapor is insensitive to the incident RF E-field. Generally, the absorption of the vapor decreases with increasing $\Omega_\mathrm{C}$. As $\Omega_\mathrm{C}$ increases, larger $\Omega_\mathrm{RF}$ is needed to spectrally split the EIT peaks, resulting in an extended AR and NL, due to power broadening.

The relative magnitudes of the coherences over the range of $\Omega_\mathrm{RF}$ plotted in \SubFigRef{Fig:NonThermalAbsorptionCoefficientRelRec}{a} suggest several approximations that can be insightful. At the upper end of the $\Omega_\mathrm{RF}$ range, $\Omega_\mathrm{RF} \sim \SIangfreq{1}{\mega \hertz}$, for $\Omega_\mathrm{P}= \SIangfreq{100}{\kilo \hertz}$ and $\Omega_\mathrm{C}= \SIangfreq{10}{\mega \hertz}$, the coupling coherence is comparable to the coherence on the RF transition. In this region, $C_\mathrm{C}= 32 \simeq C_\mathrm{RF}= 30 \gg C_\mathrm{P} = 0.02$. In this region, as well as for smaller $\Omega_\mathrm{RF}$, the weak probe approximation is well satisfied. At the lower end of the range, $\Omega_\mathrm{RF} < 2 \pi \times \SI{200}{\kilo\hertz}$, an additional approximation can be implemented since $C_\mathrm{C} \gg C_\mathrm{RF} \gg C_\mathrm{P}$. Here, the coupling laser coherence dominates and the atom can be viewed as being dressed by the coupling field alone. At this point, the dressed atom picture becomes clear. The probe laser coherence is weak in all cases due to the large scattering rate of the probe transition, $\gamma_2$, and our assumption of a small probe laser Rabi frequency. 

In the weak probe approximation, the resonant absorption coefficient is found to be
\begin{equation}
\alpha = A k_\mathrm{P} \dfrac{1}{1+\dfrac{C^2_\mathrm{C}}{1+C^2_\mathrm{RF}}} = A  k_\mathrm{P} L_0  S_0. 
\label{Eq:NonThermal_WeakProbe}
\end{equation}
$S_0 = 1/(1+ C^2_\mathrm{C}/(1+C^2_\mathrm{RF}))$ is the single atom signal, while $L_0= 1$. Plots of \EqnRef{Eq:NonThermal_WeakProbe} are shown as dashed curves in \SubFigRef{Fig:NonThermalAbsorptionCoefficientRelRec}{a} for each $\Omega_\mathrm{C}$ as a function of $\Omega_\mathrm{RF}$. The dashed curves are seen to approximate the full numerical calculation well. Taking the relative error as the difference between the full density matrix calculation and the approximation in \EqnRef{Eq:NonThermal_WeakProbe} normalized by the full density matrix calculation,  the relative error remains $<3\%$ for all $\Omega_\mathrm{C}$ shown with the error decreasing as $\Omega_\mathrm{RF}$ becomes smaller. The error as a function of $\Omega_\mathrm{RF}$ is shown in the inset. Note that the weak probe approximation works well in the NL.

When $C_\mathrm{C} \gg C_\mathrm{RF} \gg C_\mathrm{P}$, $C_\mathrm{C}$ dominates the atomic dynamics. Taking $C_\mathrm{C}\gg C_\mathrm{RF}$,
\begin{equation}
    \alpha = A k_\mathrm{P} \frac{1+C^2_\mathrm{RF} }{C^2_\mathrm{C}} = \alpha_\mathrm{EIT} +\alpha_\mathrm{RF}.
    \label{Eq:NonThermal_StrongCoupling}
\end{equation}
Under the large $C_\mathrm{C}$ approximation, $\alpha$ at small $\Omega_\mathrm{RF}$ separates into a sum of two contributions. The first term in \EqnRef{Eq:NonThermal_StrongCoupling}, $\alpha_\mathrm{EIT}=A k_\mathrm{P}/{C^2_\mathrm{C}}$, represents reduction of the absorption coefficient due to EIT.
Note that \EqnRef{Eq:NonThermal_StrongCoupling} assumes strong EIT, meaning that the coupling laser Rabi frequency dominates all dynamics, $C_\mathrm{C} \gg C_\mathrm{P}$.
The second term in \EqnRef{Eq:NonThermal_StrongCoupling} describes the RF E-field induced change of $\alpha$. The increase in $\alpha$ is governed by the ratio $C^2_\mathrm{RF}/C^2_\mathrm{C}$ and is quadratic in $\Omega_\mathrm{RF}$. It is useful to substitute our expressions for the coherences into the second term of \EqnRef{Eq:NonThermal_StrongCoupling}. If we carry-out the substitutions, we obtain,
\begin{equation}
\alpha_\mathrm{RF}= A k_\mathrm{P} \frac{C^2_\mathrm{RF}}{C^2_\mathrm{C}} = A k_\mathrm{P} \frac{\gamma_2 \Omega^2_\mathrm{RF}}{\gamma_4 \Omega^2_\mathrm{C}}.
\label{Eq:simplealpha}
\end{equation}
$\alpha_\mathrm{RF}$, in this limit, indicates that $\Omega_\mathrm{RF}/\Omega_\mathrm{C}$ quadratically enhances the probe scattering $\gamma_2$. The change in absorption of the atoms is reduced by the decay rate $\gamma_4$ of the Rydberg atoms. Collisions, transit-time broadening and other dephasing and decay processes can contribute to reducing $\alpha_\mathrm{RF}$ just as $\gamma_4$ does in \EqnRef{Eq:simplealpha}. Such effects need to be mitigated for ideal detection of the RF E-field, i.e., the case where spontaneous emission dominates. \EqnRefCap{Eq:simplealpha} reveals the parameters that need to be considered to maximize sensitivity in Rydberg atom-based sensors using all-optical detection. The quadratic expansion of $\alpha$ obtained from \EqnRef{Eq:NonThermal_StrongCoupling} is shown  in \SubFigRef{Fig:NonThermalAbsorptionCoefficientRelRec}{a} for the same $\Omega_\mathrm{C}$ as for \EqnRef{Eq:NonThermal_WeakProbe}. For small $\Omega_\mathrm{RF}\leq \SIangfreq{500}{\kilo\hertz}$, the approximation agrees well with the full calculation. The relative errors are less than $2.5\%$ in the inset of \SubFigRef{Fig:NonThermalAbsorptionCoefficientRelRec}{a}. \SubFigRefCap{Fig:NonThermalAbsorptionCoefficientRelRec}{b} shows $R_\mathrm{L}$ as a function of $\Omega_\mathrm{RF}$. For each respective $\Omega_\mathrm{C}$, $R_\mathrm{L}$ approaches the AT regime for large $\Omega_\mathrm{RF}$. For large $\Omega_\mathrm{C}$, the magnitude of $R_\mathrm{L}$ is increased, making the system more sensitive over the AR.

\subsection{Two-Photon Wave Vector Matched Limit}

We consider a hypothetical excitation scheme where the probe and coupling transitions are of the same wavelength. Consequently, the two-photon Doppler shifts cancel out, as $k_\mathrm{eff} = 0$. The single photon Doppler shifts of the probe laser transition must still be retained in the theory. Assuming $k_\mathrm{eff}=0$ leads to a narrow spectral line width, comparable to $T=0$, but the $k_\mathrm{eff}=0$ case is different because the atoms have finite velocity and $k_\mathrm{p}\neq 0$. Although this case seems contrived, it is relevant to an interpretation of how atom-based RF sensors work, providing insight into the role of Doppler shifts, and is closely related to the three-photon co-linear approach found in \cite{Kuebler2018,Bohaichuk2023threephoton}, especially if the intermediate state of the two transitions effectively serving as the coupling transition is adiabatically eliminated.

The velocity dependent absorption coefficient in the limit of $k_\mathrm{eff}= 0$ simplifies to
\begin{equation}
\alpha (v) =A k_\mathrm{P} \, \Im\left( \dfrac{i}{2i \kappa_\mathrm{P}v + S^{-1}_0}  \right).
\label{Eq:DeltaKzeroLimit}
\end{equation}
$\kappa_\mathrm{P} = k_\mathrm{P}/\gamma_2$ is the normalized wave vector, in analogy to the normalized detunings $\delta_i$, such that $\kappa_\mathrm{P}\bar{v}$ gives the Doppler half width in units of the probe transition's natural linewidth. For cesium, at room temperature, $\kappa_\mathrm{P}\bar{v} \approx 100$. $\alpha(v)$ takes the form of a Lorentzian in velocity space  with a width that is dependent on $\Omega_\mathrm{C}$ and $\Omega_\mathrm{RF}$, contained in $S_0$. 
The width of $\alpha(v)$ is comparable to the velocity distribution at room temperature, which is approximately $\SI{500}{\meter \per \second}$. The thermally weighted absorption coefficient takes the shape of a Voigt distribution in velocity space and has an overall width that is given by the velocity width of the product of $\alpha(v)$ and $P(v)$. 

\EqnRefCap{Eq:DeltaKzeroLimit} allows for an analytic integration over all velocity classes of the atoms in the thermal sample. 
Integration over the atomic velocity yields the thermally averaged absorption coefficient, $\alpha$,
\begin{align}
\alpha &= \int_{-\infty}^{\infty} P(v)\alpha(v) \, \mathrm{d}v \nonumber \\
 &= \frac{A k_\mathrm{P} }{\sqrt{\pi} \bar{v} } \int_{-\infty}^{\infty} \Im \left(  \dfrac{ i }{2i\kappa_\mathrm{P}v + S^{-1}_0} \right)  \, \mathrm{e}^{-\dfrac{ v^2}{\bar{v}^2} } \, \mathrm{d}v   \nonumber \\ 
&= \sqrt{\dfrac{\pi}{2}}\frac{ A k_\mathrm{P}  }{2\kappa_\mathrm{P}\bar{v} } \exp\left(\frac{S_0^{-1}}{2\kappa_\mathrm{P} \bar{v}} \right)^2 \Bigg(1-\erf\left( \frac{S_0^{-1}}{2\kappa_\mathrm{P} \bar{v}} \right)\Bigg) .
\label{Eq:DeltaKzeroLimitMaxwell}
\end{align}
In \EqnRef{Eq:DeltaKzeroLimitMaxwell}, the absorption of the probe laser is weakened by $2 \kappa_\mathrm{P}\bar{v}$ in a Doppler broadened medium. The ratio $\gamma_2/2 k_\mathrm{P}\bar{v}$ can be interpreted as the fraction of atoms in the thermal gas that are addressed by the probe laser.
The single-atom on resonance contribution to the absorption $S_0^{-1}=1+C^2_\mathrm{C}/(1+C^2_\mathrm{RF})$ is also rescaled by the same factor. The signal term is rescaled because the Doppler broadening also affects the two photon dephasing.

In most experiments, $S_0^{-1}/(2\kappa_\mathrm{P}\bar{v})$ is small. For $\Omega_\mathrm{C}= \SIangfreq{20}{\mega \hertz}$ and $\Omega_\mathrm{RF}=\SIangfreq{1}{\mega \hertz}$, $S_0^{-1}/(2\kappa_\mathrm{P}\bar{v})\approx 0.05$. $S_0^{-1}/(2\kappa_\mathrm{P}\bar{v})$ can be significantly smaller for smaller $\Omega_\mathrm{C}$ and larger $\Omega_\mathrm{RF}$. 
For small $S_0^{-1}/(2\kappa_\mathrm{P}\bar{v})$, $\alpha$ can be linearized,
\begin{equation}
    \alpha
    = A k_\mathrm{P} \frac{\sqrt[4]{2}}{2\kappa_\mathrm{P}\bar{v} } \left( \frac{\sqrt{\pi}}{2\sqrt[4]{2}} - \frac{\sqrt[4]{2}}{2\kappa_\mathrm{P} \bar{v}}S^{-1}_0 \right).
    \label{Eq:DeltaKzeroLinApprox}
\end{equation}
\EqnRefCap{Eq:DeltaKzeroLinApprox} is reminiscent of the result obtained for $T=0$ under the weak probe approximation.
Linearizing $\alpha\propto1/S_0^{-1}$ for $T=0$ around $S_0^{-1}=1$, \EqnRef{Eq:NonThermal_WeakProbe}, leads to $\alpha = A k_\mathrm{P} (2-S^{-1}_0)$, since $S_0 \simeq 2- S^{-1}_0$.
Doing the same around $S_0^{-1}=2\kappa_\mathrm{P}\bar{v}/\sqrt[4]{2}$, \EqnRef{Eq:NonThermal_WeakProbe}, leads to ${\alpha =A k_\mathrm{P} \frac{\sqrt[4]{2}}{2\kappa_\mathrm{P}\bar{v}}\left(2-\frac{\sqrt[4]{2}}{2\kappa_\mathrm{P} \bar{v}} S^{-1}_0 \right)}$, differing only in an offset from \EqnRef{Eq:DeltaKzeroLinApprox}. The latter expansion point is more realistic for the strong coupling regime. The strong coupling result strengthens the interpretation that the signal is mainly caused by contributions of velocity classes around the mean thermal velocity. The $\sqrt[4]{2}$ comes from the fact that the Voigt profile is broader than the original Lorentzian in velocity space.
Again, as in \EqnRef{Eq:DeltaKzeroLimitMaxwell}, the coefficient $A \sqrt[4]{2}/2 \kappa_\mathrm{P}
\bar{v}$ can be interpreted as the number of atoms in the probe laser mode volume that are available for interaction with the laser fields.
For large $C_\mathrm{C}$, $C^2_\mathrm{C} \rightarrow C^2_\mathrm{C} \cdot 2 \kappa_\mathrm{P} \bar{v}/\sqrt[4]{2}$, \EqnRef{Eq:DeltaKzeroLimitMaxwell} and \EqnRef{Eq:DeltaKzeroLinApprox} show that the Doppler broadening reduces both the overall magnitude of the absorption coefficient and the effect of the modifying signal on the absorption coefficient. The influence of the thermal motion of the atoms can be largely incorporated by considering the mean Doppler broadening.

\subsection{Two-Photon Finite Wave Vector Mismatch at Finite Temperature}\label{Sec:Thermal}

The experiments done in atom-based RF E-field sensing to date all have finite wave vector mismatch and are performed in a thermal atomic vapor. The probe laser and 2-photon Doppler shifts modify the optical response leading to a more complex optically addressable velocity distribution. Finite Doppler shifts are a basic part of RF E-field, vapor cell sensors. 

\EqnRefCap{Eq:OpticalResponse4Level} provides a general expression for the optical response $\rho_{12}$. Substituting the Doppler shifts according to \EqnRef{Eq:Thermal_DopplerShift} into \EqnRef{Eq:OpticalResponse4Level} and subsequently into \EqnRef{Eq:GeneralAlphaCalcThermal} leads to an integral expression for $\alpha$,
\begin{equation}
\alpha = A k_\mathrm{P}\, \Im \int_{-\infty}^\infty  \frac{1 }{2\delta_\mathrm{P}(v)-i}  \dfrac{1}{1-\dfrac{\dfrac{C^2_\mathrm{C}}{(2\delta_\mathrm{P}(v)-i)( 2\delta_\mathrm{C}(v)-i)}}{1- \dfrac{C^2_\mathrm{RF}}{( 2\delta_\mathrm{C}(v)-i)( 2\delta_\mathrm{RF}(v)-i)}}}  P(v) \, \mathrm{d}v.
\label{Eq:Thermal_AbsorptionCoefficient}
\end{equation}
The finite wave vector mismatch of the excitation lasers and non-zero temperature of the atomic gas complicate the description of the absorption constant and lead to an expression that cannot be analytically integrated. In \EqnRef{Eq:Thermal_AbsorptionCoefficient}, the velocity dependence of the normalized detunings $\delta_i(v)$ is highlighted.
\begin{figure*}
    \centering
    \includestandalone[width = \textwidth]{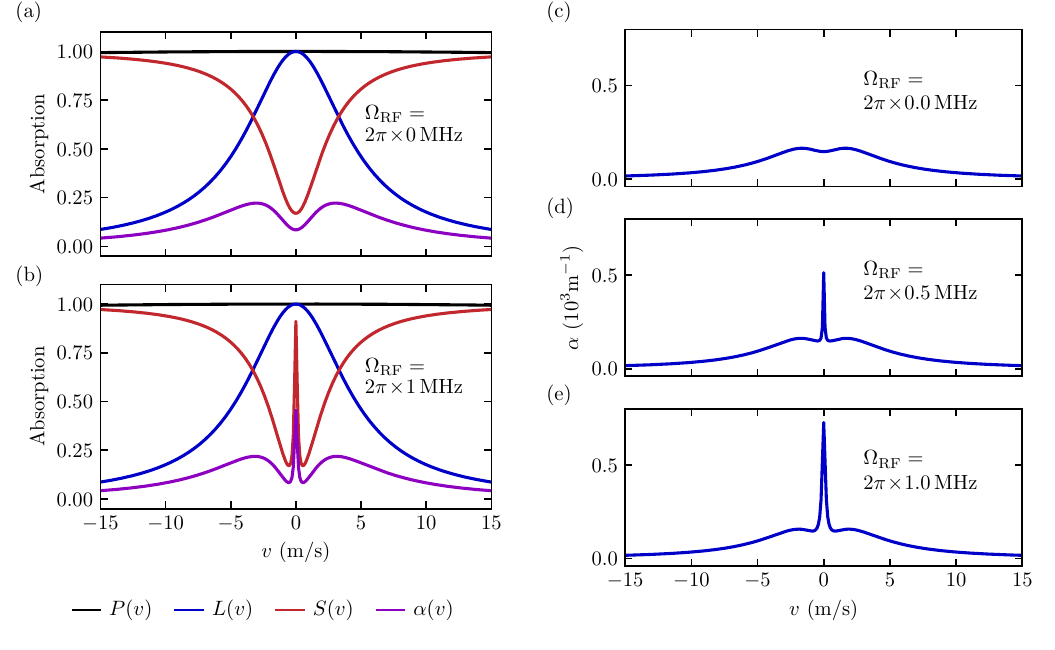}
    \caption{ Illustration of the different contributions $L(v)$, $S(v)$ and $P(v)$ to $\alpha(v)$ as a function of the atomic velocity, $v$. (a) shows the case where $\Omega_\mathrm{RF}=0$. The product of $L(v)$ and $S(v)$ leads to a broad double-peak structure of $\alpha(v)$ (solid purple curve). (b) Illustration at $\Omega_\mathrm{RF} = \SIangfreq{1}{\mega \hertz}$. The RF E-field leads to a narrow absorption feature around $v=0$. (c)-(e) $\alpha(v)$ as a function of $v$ for various $\Omega_\mathrm{RF}$. The peak height increases with increasing $\Omega_\mathrm{RF}$. The width of the broad double-peak feature remains constant. $\Omega_\mathrm{C} = \SIangfreq{5}{\mega \hertz}$ and $\Omega_\mathrm{P}=\SIangfreq{100}{\kilo \hertz} $ for these plots.}
    \label{Fig:SignalContributions}
\end{figure*}

\SubFigRefCap{Fig:SignalContributions}{a-b} shows the velocity distributions for $L(v)$, $S(v)$ and $P(v)$ as defined by \EqnRef{Eq:GeneralAlphaCalcThermal}. The curves are normalized to compare their line shapes. Over the extent of significant absorption, the velocity distribution, $P(v)$, is almost constant and the largest addressed velocity class is small compared to $\bar{v}$. The probe transition Lorentzian $L(v)$ forms an envelope for $\alpha(v)$. The width of $L(v)$ is proportional to $\gamma_2 /k_\mathrm{P}\approx \SI{6}{\meter \per \second}$. The width of $L(v)$, $\gamma_2/k_\mathrm{P}$, characterizes the velocity classes of the atomic sample that absorb probe light. 
In the case of $\Omega_\mathrm{RF}=0$, \SubFigRef{Fig:SignalContributions}{a}, $S(v)$ strongly modifies the probe Lorentzian around $v = 0$, leading to a double-peak structure for $\alpha(v)$, i.e., EIT. The width of $S(v)$ is proportional to $\sqrt{\gamma_2 \gamma_3}/|k_\mathrm{eff}|\approx \SI{3}{\meter \per \second }$. The sign and magnitude of $k_\mathrm{eff}$ plays a crucial role in determining the amplitude and width of $S(v)$ as described in \SecRef{Sec:Doppler effect}. $S(v)$ can be power broadened, depending on the strength of $\Omega_\mathrm{C}$ and $\Omega_\mathrm{RF}$. 
The broad double-peak feature of $S(v)$ in \SubFigRef{Fig:SignalContributions}{b} corresponds to the absorption at $\Omega_\mathrm{RF}=0$ and the narrow feature corresponds to the increased absorption induced by $\Omega_\mathrm{RF}$. The width of the narrow feature is $\sqrt{\gamma_3 \gamma_4}/2|k_\mathrm{eff}|\approx \SI{0.1}{\meter \per \second}$. The RF E-field only addresses atoms that are almost at rest. The central peak amplitude increases and its width broadens as $\Omega_\mathrm{RF}$ increases. \SubFigRefCap{Fig:SignalContributions}{c-e} shows $\alpha(v)$ as a function of velocity, where all terms are to scale. The absorption coefficient is calculated for different $\Omega_\mathrm{RF}$. The RF feature is the dominant contribution to $\alpha(v)$ when $\Omega_\mathrm{RF} \neq 0$. The broad double peaked structure does not change as a function of $\Omega_\mathrm{RF}$.

\begin{figure*}
    \centering
    \includestandalone[width = \textwidth]{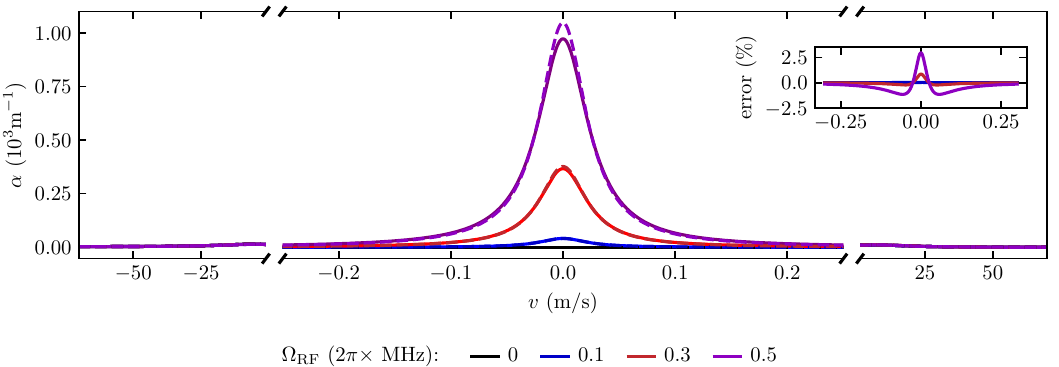}
    \caption{Comparison of the velocity distribution of the quadratic expansion (dashed lines) from \EqnRef{Eq:Thermal_QuadraticExpansion} and full numerical simulation (solid lines) for different $\Omega_\mathrm{RF}$.  The inset shows the relative deviation of the quadratic expansion from the full numerical simulation. $\Omega_\mathrm{P} = \SIangfreq{100}{\kilo \hertz}$ and $\Omega_\mathrm{C}= \SIangfreq{10}{\mega \hertz}$.}
    \label{Fig:Thermal_VelocityDistribution_QuadExp_Comp}
\end{figure*}

For weak $\Omega_\mathrm{RF}$, the two velocity space features can be separated. The separation is achieved by expanding the expression for $\alpha(v)$ in \EqnRef{Eq:GeneralAlphaCalcThermal} quadratically in $\Omega_\mathrm{RF}$. The approach is motivated by the agreement of the quadratic expansion at $T=0$ with the full density matrix calculation, see \SecRef{Sec:Teq0}. The quadratic expansion of \EqnRef{Eq:Thermal_AbsorptionCoefficient} yields 
\begin{widetext}
\begin{align}
\alpha(v)&= A k_\mathrm{P} \dfrac{1+4\delta^2_\mathrm{C}+C^2_\mathrm{C} } { \left( C^2_\mathrm{C}+4 \delta_\mathrm{P} \delta_\mathrm{C} \right)^2 +2 C^2_\mathrm{C} + 4 \delta^2_\mathrm{P} + 4\delta^2_\mathrm{C} +1 } \nonumber \\ 
&+ \frac{A k_\mathrm{P} C^2_\mathrm{C} C^2_\mathrm{RF}}{(1+4\delta^2_\mathrm{RF})} \bigg( \frac{\left( C^2_\mathrm{C}+4 \delta_\mathrm{P} \delta_\mathrm{C} \right)^2 +2 C^2_\mathrm{C} - 4 \delta^2_\mathrm{P} - 4\delta^2_\mathrm{C} +1 }{\left( C^2_\mathrm{C}+4 \delta_\mathrm{P} \delta_\mathrm{C} \right)^2 +2 C^2_\mathrm{C} + 4 \delta^2_\mathrm{P} + 4\delta^2_\mathrm{C} +1 }  \nonumber \\
&+ \frac{-8C^2_\mathrm{C}\delta_\mathrm{RF}(\delta_\mathrm{P} + \delta_\mathrm{C})- 8 \delta_\mathrm{RF} ( \delta_\mathrm{P}+ \delta_\mathrm{C} ) - 16 \delta_\mathrm{P} \delta_\mathrm{C} +32 \delta_\mathrm{P}\delta_\mathrm{C}\delta_\mathrm{RF}(\delta_\mathrm{P}+\delta_\mathrm{C})}{\left(\left( C^2_\mathrm{C}+4 \delta_\mathrm{P} \delta_\mathrm{C} \right)^2 +2 C^2_\mathrm{C} + 4 \delta^2_\mathrm{P} + 4\delta^2_\mathrm{C} +1 \right)^2} \bigg).
\label{Eq:Thermal_QuadraticExpansion}
\end{align} 
\end{widetext}
In \FigRef{Fig:Thermal_VelocityDistribution_QuadExp_Comp}, the numerical calculation of $\alpha(v)$ is compared to \EqnRef{Eq:Thermal_QuadraticExpansion}, dashed lines. \FigRefCap{Fig:Thermal_VelocityDistribution_QuadExp_Comp} shows the velocity distribution of the resonant absorption coefficient over a broad range of velocity classes for several $\Omega_\mathrm{RF}$.
The broad envelope and the double peak structure in the velocity distribution of $\alpha$ are well reproduced by the quadratic expansion. 

The narrow feature, around $v=0$, of $\alpha$ is emphasized in the center section of \FigRef{Fig:Thermal_VelocityDistribution_QuadExp_Comp}. The height and width of the narrow feature are well captured by the approximation at small $\Omega_\mathrm{RF}$. At large $\Omega_\mathrm{RF}$, the quadratic expansion overshoots the full calculation. The overshoot is expected because the quadratic expansion continues to increase as $\Omega_\mathrm{RF}$ grows, while the actual absorption converges to the background absorption for large $\Omega_\mathrm{RF}$ as AT splitting starts to occur. The relative errors for the narrow feature are shown in the inset of \FigRef{Fig:Thermal_VelocityDistribution_QuadExp_Comp}. As the general shape of the narrow feature is well replicated by the quadratic expansion, the quadratic expansion can be used to obtain the RF E-field in the AR. 

The quadratic expansion of $\alpha$, \EqnRef{Eq:Thermal_QuadraticExpansion}, consists of two terms that can be independently integrated over the atomic velocity distribution. To make the integration analytically possible, the system is assumed to be in the strong coupling limit. Within this limit, the coupling laser Rabi frequency dominates over the natural decay rates of the Rydberg transition and the respective Doppler shifts,
\begin{equation}
\Omega^2_\mathrm{C} \gg \gamma_2 \gamma_3, \hspace{0.75cm } \Omega_\mathrm{C} \gg k_\mathrm{P}v, \hspace{0.75cm} \Omega_\mathrm{C}\gg k_\mathrm{C}v.
\label{Eq:Thermal_Approximations}
\end{equation}
With the approximations from eqs.~(\ref{Eq:Thermal_Approximations}), the expression for the absorption coefficient, $\alpha(v)$, simplifies to
\begin{equation}
\alpha(v) = \frac{A k_\mathrm{P} }{4\delta^2_\mathrm{P}} \dfrac{1+\dfrac{C^2_\mathrm{C}}{4\delta^2_\mathrm{C}}}{1+\dfrac{\left( C^4_\mathrm{C}-8 C^2_\mathrm{C}\delta_\mathrm{P} \delta_\mathrm{C} \right)}{16\delta^2_\mathrm{P} \delta^2_\mathrm{C}}}  \left( 1+ \frac{C^2_\mathrm{RF}}{1+4\delta^2_\mathrm{RF}}\right).
\label{Eq:Thermal_Alpha_Approx}
\end{equation}
The first term in \EqnRef{Eq:Thermal_Alpha_Approx} is the EIT background term. The second term in \EqnRef{Eq:Thermal_Alpha_Approx} describes the absorption change induced by the RF E-field. The final expression for $\alpha(v)$ under these approximations is integrable and yields the Doppler-averaged absorption coefficient, $\alpha$,
\begin{widetext}
\begin{align}
\alpha &=  \frac{A k_\mathrm{P}}{ \sqrt{\pi}\bar{v}} \int_{-\infty}^{\infty}  \frac{ (C^2_\mathrm{C} + 4\delta^2_\mathrm{C} ) }{\left( C^2_\mathrm{C}-4\delta_\mathrm{P} \delta_\mathrm{C} \right)^2} \left( 1+\frac{C^2_\mathrm{RF} }{1+4\delta^2_\mathrm{RF}} \right)\, \mathrm{d}v \nonumber\\ 
&=\frac{ A k_\mathrm{P} \sqrt{\pi}}{\kappa_\mathrm{P}\bar{v}} \left( \frac{ \kappa_\mathrm{C}- \kappa_\mathrm{P}}{4 C_\mathrm{C} \sqrt{\kappa_\mathrm{P}\kappa_\mathrm{C}}} + \frac{C^2_\mathrm{RF} (- \sqrt{ \kappa_\mathrm{P}\kappa_\mathrm{C} } (\kappa_\mathrm{C}-\kappa_\mathrm{P} )  + 2 \kappa_\mathrm{P}\kappa_\mathrm{RF} C_\mathrm{C} ) }{ 4 C_\mathrm{C} (\sqrt{\kappa_\mathrm{P} \kappa_\mathrm{C}} +\kappa_\mathrm{RF} C_\mathrm{C} )^2 }\right), 
\label{Eq:Thermal_Integration_FullTerm}
\end{align}
\end{widetext}
where $\kappa_\mathrm{C}= |k_\mathrm{eff}|/\gamma_3$ and $\kappa_\mathrm{RF}= |k_\mathrm{eff}|/\gamma_4$ are introduced as analogous parameters to $\kappa_\mathrm{P}$, \EqnRef{Eq:DeltaKzeroLimit}. Note that we took the negative sign of $k_\mathrm{eff}$ into account in \EqnRef{Eq:Thermal_Integration_FullTerm}. The normalization factor $\bar{v}$ comes from the velocity distribution. 
In the strong coupling approximation, we can estimate the size of the quantities appearing in the second term of \EqnRef{Eq:Thermal_Integration_FullTerm} to simplify the expression.  For normal experimental conditions in the AR,  $\sqrt{\kappa_\mathrm{P} \kappa_\mathrm{C}} \approx \SI{3}{\second \per \meter }$ while $ \kappa_\mathrm{RF} C_\mathrm{C} \approx \SI{800}{\second \per \meter }$. Using these estimates, we can approximate \EqnRef{Eq:Thermal_Integration_FullTerm},
\begin{equation}
\alpha \simeq \frac{ A k_\mathrm{P} \sqrt{\pi}}{\kappa_\mathrm{P}\bar{v}} \left( \frac{\kappa_\mathrm{C}- \kappa_\mathrm{P}}{4 C_\mathrm{C} \sqrt{\kappa_\mathrm{P}\kappa_\mathrm{C}}} + \frac{ \kappa_\mathrm{P}C^2_\mathrm{RF}}{2 \kappa_\mathrm{RF}C^2_\mathrm{C} } \right).
\label{Eq:Thermal_Integration}
\end{equation}
\EqnRefCap{Eq:Thermal_Integration} shows similar behavior to the quadratic expression for $\alpha$ in \EqnRef{Eq:NonThermal_StrongCoupling} and \EqnRef{Eq:DeltaKzeroLinApprox}. Both expressions for $\alpha$ contain a term that produces the EIT transmission for a specific $C_\mathrm{C}$ at $\Omega_\mathrm{RF}=0$. The transmission is reduced by the Doppler shift contributions for finite wave vector mismatch and finite temperature as in the case of $k_\mathrm{eff}=0$. 
The effect of the RF E-field is manifested in the ratio $C^2_\mathrm{RF}/C^2_\mathrm{C}$. The increase in absorption due to the second term is reduced by the ratio of the probe transition and upper Rydberg state dephasing rates and the ratio of $k_\mathrm{P}$ and $k_\mathrm{eff}$. The signal reduction is more similar to the $T=0$ than the $k_\mathrm{eff}=0$ case. The difference arises from the fact that the RF signal is due to a narrow range of velocities around 0, similar to $T=0$.

\begin{figure*}
\centering
\includestandalone[width = \textwidth]{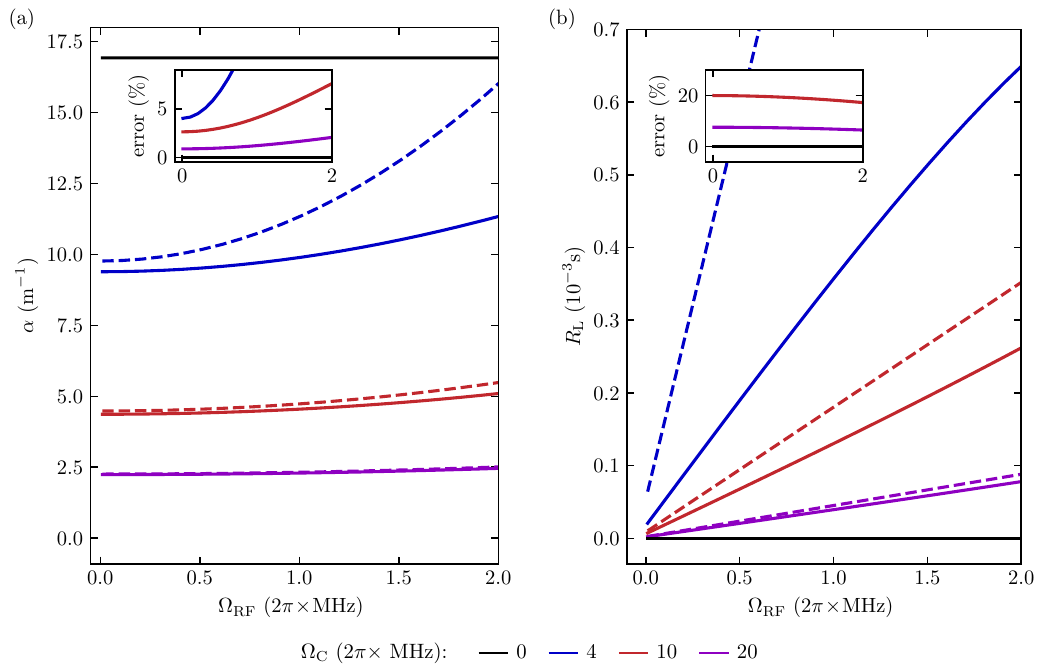}
\caption{(a) Absorption coefficient, $\alpha$, as a function of $\Omega_\mathrm{RF}$ for various $\Omega_\mathrm{C}$ with Doppler averaging included. Generally, $\alpha$ behaves similarly to the $T=0$ case, see \SubFigRef{Fig:NonThermalAbsorptionCoefficientRelRec}{a}. Due to the additional Doppler broadening, the AR is extended to larger $\Omega_\mathrm{RF}$. The dashed lines correspond to the approximate expression, see \EqnRef{Eq:Thermal_Integration}. In the strong coupling limit, the approximation agrees with the full calculation within a relative error of $<5\%$. (b) The responsitivity, $R_\mathrm{L}$ as a function of $\Omega_\mathrm{RF}$. The probe Rabi frequency is $\Omega_\mathrm{P} = \SIangfreq{100}{\kilo \hertz}$.
}
\label{Fig:Thermal_AbsCoeffRelAbsRecalcRF}
\end{figure*}

In \SubFigRef{Fig:Thermal_AbsCoeffRelAbsRecalcRF}{a}, $\alpha$ is shown as a function of $\Omega_\mathrm{RF}$ for several typical $\Omega_\mathrm{C}$. $\alpha$ behaves similarly to the $T=0$ case. 
For increasing $\Omega_\mathrm{C}$, $\alpha (\Omega_\mathrm{RF}=0)$ decreases and the presence of small RF E-fields leads to enhanced absorption in the system, comparable to $T=0$. The approximate expression given by \EqnRef{Eq:Thermal_Integration} is shown as dashed curves for the respective $\Omega_\mathrm{C}$. The approximation approaches the numerical calculation in the limit of large $\Omega_\mathrm{C}$. The relative errors are shown as the inset in \FigRef{Fig:Thermal_AbsCoeffRelAbsRecalcRF}.

\section*{Discussion}

In the weak probe limit, where the probe laser Rabi frequency is small compared to the natural decay rate of the transition, useful analytical expressions for the absorption coefficient can be found and can be accurately applied from the AR through the NL. The weak probe approximation is remarkably accurate over the AR and NL, particularly for $T=0$.  In ladder-like excitation schemes, the absorption coefficient can be expressed as a product of the probe transition Lorentzian, $L$, and a modifying signal term, $S$. The probe transition Lorentzian behaves as the corresponding two-level system. The signal term modifies the photon scattering rate on the probe transition and thereby the absorption of the atoms. $S$ depends on the coherent driving fields of the other atomic levels that are involved in the multi-level interference. The RF E-field strength is carried to the probe laser absorption signal via the coupling transition coherence. The modifying signal takes the form of a continued fraction, where the roles of the auxiliary atomic resonances and fields can clearly be identified. We made a large $\Omega_\mathrm{C}$ approximation in addition to the weak probe approximation. In the large $\Omega_\mathrm{C}$ approximation the analytic expressions are amenable to clear physical interpretations. We used the coherences of the atomic transitions in the expression for the absorption coefficient. The coherence parameter is a measure of the number of Rabi oscillations that occur during the decay time of the transition. In the case where the incident RF field is small, both the coherences of the probe and the RF sensing transition are small. The situation lends itself to an interpretation where the atom is dressed by the coupling laser field to create a new 'atom' whose absorption changes in a controlled way in the presence of an RF E-field.

The analytic expressions for the absorption coefficients under different conditions can be used to recalculate the RF E-field strength in the AR and NL. With knowledge of the natural decay rates as well as the coupling Rabi frequency, the RF E-field can be recalculated.
Relative errors are presented, referenced to the full numerical calculations.

For $T=0$, all atoms are at rest and resonantly addressed by the electromagnetic fields. The $T=0$ limit illustrates that each atom behaves as an individual antenna. We identified expressions for the single-atom probe transition Lorentzian and modifying signal term. The absorption of the probe beam increases in proportion to $\Omega^2_\mathrm{RF}/\Omega^2_\mathrm{C}$ at small $\Omega_\mathrm{RF}$. 
The modification of the probe transition scattering rate is reduced by the spontaneous decay rate of the Rydberg state coupled to the system by the RF E-field. The decay rate of the Rydberg state, $\gamma_4$, coupled to the EIT system by the RF E-field is the fundamental limiting factor for the change in absorption. Decoherence of the Rydberg transition decreases sensitivity. The dephasing and decay rates can comprise contributions from other phenomena like collisions and transit time broadening. The decay and dephasing mechanisms need to be mitigated to achieve the best sensitivity to the RF E-field. 

We included thermal motion of the atoms in a hypothetical excitation scheme, where the wavelength of the excitation lasers are the same and the two-photon detuning cancels out in a counter-propagating geometry, $k_\mathrm{eff}=k_\mathrm{P}+k_\mathrm{C}=0$.
The wave vector matching leads to a narrow spectral linewidth.
For $k_\mathrm{eff}=0$, $\alpha$ as a function of velocity has a Lorentzian profile with a width that is given by the single-atom signal. The width of the velocity distribution is comparable to the mean velocity of the thermal gas. The Lorentzian shaped velocity distribution of $\alpha$ and the Gaussian velocity distribution of the atoms lead to a Voigt distribution in velocity space. The resulting absorption coefficient can be integrated analytically over the atomic velocity. For weak $\Omega_\mathrm{RF}$, the thermally averaged expression for $\alpha$ can be expanded linearly in the single-atom signal. In expanded form, $\alpha$ has a similar form to the result obtained for $T=0$. However, the mean Doppler broadening, $2k_\mathrm{P}\bar{v}/\gamma_\mathrm{2}$, reduces the fraction of atoms addressed by the probe laser as well as the change in photon scattering on the probe laser transition. The mean Doppler broadening reduces the absorption signal amplitude compared to $T=0$. Due to the mean Doppler broadening, the absorption for $k_\mathrm{eff}=0$ is less than for $T=0$ given equal numbers of atoms. Although this case is artificial, it reveals the role of Doppler broadening and is closely approximated by the three-photon, co-linear method found in~\cite{Kuebler2018,Bohaichuk2023threephoton}.

Finite wave vector mismatch and a thermal atomic gas leads to a more complicated structure in the velocity distribution of $\alpha$. The magnitude and sign of $k_\mathrm{eff}$ plays a crucial role in determining the signal height and spectral width of the absorption spectrum. The velocity distribution of $\alpha$ consists of two distinct features.
For $\Omega_\mathrm{RF}=0$, a broad double peak structure emerges from the product of $L(v) S(v)$. The width of the feature strongly depends on $\Omega_\mathrm{C}$, but remains small compared to $\bar{v}$.
For finite $\Omega_\mathrm{RF}$, an additional narrow velocity feature emerges around $v=0$ that dominates the absorption amplitude for larger $\Omega_\mathrm{RF}$ and is the important feature for RF E-field sensing. Only atoms moving with small velocities can sense a change of $\Omega_\mathrm{RF}$.
For weak incident $\Omega_\mathrm{RF}$, the two features can be treated independently. The superposition of the terms describing the broad and narrow features in velocity space separate the absorption at $\Omega_\mathrm{RF}=0$ from the absorption increase due to finite $\Omega_\mathrm{RF}$. 
In the strong coupling limit, where $\Omega_\mathrm{C}$ is considered to dominate over the natural decay rates of the coupling laser transition and Doppler detunings, $\alpha(v)$ can be analytically integrated. The resulting thermally averaged absorption coefficient shows similar behavior to $T=0$ and $k_\mathrm{eff}=0$ with an amplitude regime extended to larger $\Omega_\mathrm{RF}$ since the absorption profile is broadened by the larger Doppler effect. The increase in absorption due to $\Omega_\mathrm{RF}$ is proportional to $C^2_\mathrm{RF}/C^2_\mathrm{C}$ analogous to $T=0$ and $k_\mathrm{eff} =0$. However, the $T=0$ and finite wave vector mismatch with finite temperature cases are more similar because in both situations the atoms with velocities around $v=0$ dominate the signal.

\section{Sensitivity}

Evaluating sensitivity in the amplitude regime is important for determining the lowest detectable RF E-field.
In many cases, Rydberg atom-based sensors are limited by the shot-noise of the probe laser. The laser shot-noise is the fundamental quantum fluctuations of the probe laser field and the probabilistic light detection process. The probe laser shot noise directly depends on the number of photons arriving at the detector,
\begin{equation}
    N_\mathrm{ph}(\Omega_\mathrm{RF})=\frac{\exp(-l\alpha(\Omega_\mathrm{RF}))\,P_0\,\lambda}{h c B}=\frac{P_1(\Omega_\mathrm{RF})\,\lambda}{h c B},
\end{equation}
where $l$ is the path length of the probe laser, $c$ is the speed of light, $P_0$ is the probe laser power incident on the vapor cell, $P_1$ is the power transmitted through the vapor cell and $\lambda$ is the probe laser wavelength. The bandwidth of the measurement is denoted $B$. We define the shot-noise limited absorption coefficient for a signal-to-noise ratio of one, such that the change in number of transmitted photons due to the presence of the RF field is equal to the expected photon number fluctuation for a Poissonian distribution, i.e. a classical field produced by the probe laser,
\begin{equation}
    \sqrt{N_\mathrm{ph}(0)}=\left|N_\mathrm{ph}(\Omega_\mathrm{RF})-N_\mathrm{ph}(0)\right|.
\end{equation}
Dividing by $N_\mathrm{ph}(0)$ and assuming the overall absorption is small, results in an expression for the minimum resolvable change in absorption,
\begin{equation}
    \sqrt{\frac{hcB}{P_1(0)\lambda}} \approx (\alpha(\Omega_\mathrm{RF})-\alpha(0))l=\OD(\Omega_\mathrm{RF})-\OD(0)=\Delta\OD,
\end{equation}
where we introduce the optical depth, $\OD = \alpha l$. We also take the quantum efficiency of the detection process to be unity, since we are investigating the relative sensitivities of the excitation and read-out schemes described in the paper. We choose the $\OD$ to evaluate sensitivity because it accounts for the atom density as well as the geometric size of the laser-cesium gas interaction volume. The background absorption coefficient, $\alpha(\Omega_\mathrm{RF}=0)$ is limited by the EIT signal, usually $\exp(-\alpha(\Omega_\mathrm{RF}=0)l) \approx 0.3-0.7$. The density of Rydberg atoms is limited by collisions since these can cause dephasing of the Rydberg state.

To numerically estimate the sensitivity, we use a representative set of parameters that are achievable in Rydberg atom-based RF sensors.
We assume a probe laser beam resonant with the Cs D2 line, $\lambda = \SI{852}{\nano \meter}$. We choose laser beam diameters of $d= \SI{1}{\milli \meter}$ and a vapor cell length of $\SI{3}{\centi \meter}$, which sets the interaction region volume in the vapor cell. We choose $P_1(0) = \SI{20}{\micro\watt}$. To get realistic results, we include transit time broadening of $\SI{200}{\kilo \hertz}$, corresponding to room temperature and $d= \SI{1}{\milli \meter}$. Transit time broadening is modeled as an additional decay channel for each excited state that couples it to the ground state.
We use an orientationally averaged RF transition radial dipole moment of 
$\mu_\mathrm{RF} = \SI{2569.7}{\elementarycharge \bohrradius}$, corresponding to the 55D$_{5/2}$ to 53F$_{7/2}$ transition in cesium.
For $T=0$, we carry-out two different calculations. First, we use the same number of atoms as can be accessed in a vapor cell of the aforementioned geometry. In a second calculation, we use a magneto-optic trap (MOT) to calculate the sensitivity to obtain a more accurate comparison of sensitivity between $T=0$ and a room temperature vapor cell. Typical MOTs reach peak densities of \SI{1e10}{\per \centi \meter\cubed} with a diameter of \SI{1}{\milli \meter}. The total number of atoms in the MOT is $\approx \num{1e7}$. A thermal vapor at room temperature, $T=\SI{20}{\celsius}$, has an average density of \SI{3e10}{\per \centi \meter\cubed}~\cite{Steck2019}.
The total number of atoms in the interaction volume considered here is \num{7e8}, \num{\sim 70} times larger than for the MOT.
All quantities are quoted for a \SI{500}{\kilo\hertz} bandwidth. We chose a $\SI{500}{\kilo\hertz}$ bandwidth because it corresponds to a pulse width that can be used for communications and radar and is well-suited to the atomic response time~\cite{Bohaichuk2022}.

\begin{figure*}
\centering
\includestandalone[width = \textwidth]{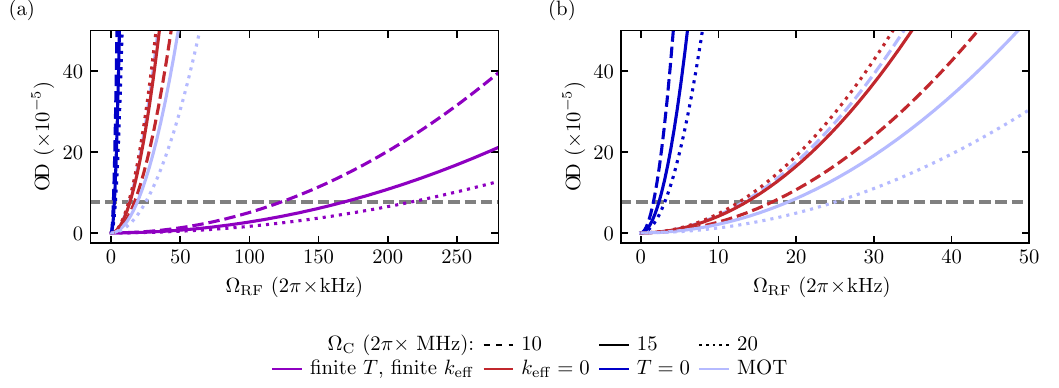}
\caption{(a) Optical densities ($\OD$) for the different cases described in the paper as a function of $\Omega_\mathrm{RF}$ for various $\Omega_\mathrm{C}$. The horizontal dashed grey line corresponds to the shot-noise limit. The intersection with the $\OD$ curves corresponds to the minimum detectable $\Omega_\mathrm{RF,min}$. (b) Zoom into the small $\Omega_\mathrm{RF}$ range. For the thermal limits an additional transit time broadening of $\SI{200}{\kilo \hertz}$ is considered. The probe Rabi frequency is $\Omega_\mathrm{P} = \SIangfreq{100}{\kilo \hertz}$.}
\label{Fig:OpticalDensityComparison}
\end{figure*}

In \FigRef{Fig:OpticalDensityComparison}, $\OD$ is compared for small RF E-fields, i.e. small $\Omega_\mathrm{RF}$ for each case described in the paper. The horizontal line corresponds to the probe laser photon shot-noise limit for detection, i.e., signal-to-noise of one. The $\OD$ are shown for several $\Omega_\mathrm{C}$ as a function of $\Omega_\mathrm{RF}$. The slope of each curve depends on $\Omega_\mathrm{C}$. In some cases, the slope of the $\OD$ decreases and in other cases increases with increasing $\Omega_{C}$. The behavior is indicative of a maximum in the slope as a function of $\Omega_{C}$. We investigated the maxima in slope numerically. For the thermal limit with finite wave vector mismatch, the slope of the $\OD$ decreases with increasing $\Omega_\mathrm{C}$ for the $\Omega_\mathrm{C}$ plotted in the figure. The maximum slope is found at $\Omega_\mathrm{C} \approx \SIangfreq{6}{\mega \hertz}$. In contrast, for $k_\mathrm{eff}=0$, the slope increases with $\Omega_\mathrm{C}$. The maximum slope is found at $\Omega_\mathrm{C} \approx \SIangfreq{24}{\mega \hertz}$. For the low temperature cases, the slope decreases with increasing $\Omega_\mathrm{C}$. The maximum slopes are found at $\Omega_\mathrm{C} \approx \SIangfreq{3}{\mega \hertz}$ (MOT) and $\Omega_\mathrm{C} \approx \SIangfreq{4}{\mega \hertz}$ ($\mathrm{T}=0$). The differences in position of the $\OD$ slope maxima with respect to $\Omega_\mathrm{C}$ explain the relative order of the curves with respect to $\Omega_\mathrm{C}$ shown in \FigRef{Fig:OpticalDensityComparison}.

The minium detectable $\Omega_\mathrm{RF,min}$ can be determined from the intersection of the $\OD$ curve and shot noise limit shown in \FigRef{Fig:OpticalDensityComparison}. With $\Omega_\mathrm{RF,min}$, the minimal detectable E-field and sensitivity can be calculated. The results are summarized in \TblRef{Tab:Sensitivities} for a bandwidth of $\SI{500}{\kilo\hertz}$. For finite temperature and wave vector mismatch, the sensitivity is one to two orders of magnitude worse than the other cases due to larger Doppler broadening associated with finite wave vector mismatch and temperature. For the smallest $\Omega_\mathrm{C} = \SIangfreq{10}{\mega \hertz}$, the sensitivity is
\SI{56}{\nano \volt \per \centi \meter \per \hertz\tothe{1/2}}. For the range of $\Omega_\mathrm{C}$, the sensitivity varies by less than a factor of two. Unsurprisingly, the best sensitivity is achieved for $T=0$,
\SI{630}{\pico \volt \per \centi \meter \per \hertz\tothe{1/2}} for $\Omega_\mathrm{C}= \SIangfreq{10}{\mega \hertz}$. All atoms occupy the same velocity class and are addressed resonantly by the light fields so the sensitivity is best. The difference between the sensitivity for $T=0$ and $k_\mathrm{eff}=0$ is less than a factor of $\sim 10$ for all $\Omega_\mathrm{C}$. For $k_\mathrm{eff}=0$, the sensitivity is 
\SI{7.5}{\nano \volt \per \centi \meter \per \hertz\tothe{1/2}} at $\Omega_\mathrm{C}= \SIangfreq{10}{\mega \hertz}$. The high sensitivity for $k_\mathrm{eff}=0$ is attributed to the narrow spectral linewidth.

Interestingly, when we compare the MOT to $k_\mathrm{eff}=0$, the vapor cell result has comparable sensitivity. Taking into account typical atom numbers in a MOT, the $\OD$ at $T =\SI{1}{\micro \kelvin }$ leads to a sensitivity of 
\SI{5.6}{\nano \volt \per \centi \meter \per \hertz\tothe{1/2}} for $\Omega_\mathrm{C}= \SIangfreq{10}{\mega \hertz}$. For $k_\mathrm{eff}=0$, the sensitivity reaches
\SI{5.2}{\nano \volt \per \centi \meter \per \hertz\tothe{1/2}} for $\Omega_\mathrm{C}= \SIangfreq{20}{\mega \hertz}$. The sensitivity for the MOT is limited by achievable atom numbers. In reality, the comparison is more favorable for the vapor cell result, since the MOTs that have been obtained on chips in vapor cells have more than an order of magnitude less atoms than our estimate. We conclude that besides the experimental overhead, cold atom systems are not the best solution for achieving high sensitivities for Rydberg atom-based RF sensors. The larger number of atoms available in vapor cells can boost the sensitivity of thermal vapor cell systems beyond what is achievable in a MOT. As a consequence, the best conditions for RF E-field sensing are currently given by an atomic vapor cell and an excitation scheme where Doppler shifts cancel out, $k_\mathrm{eff}\approx 0$. Current experiments, using a three-photon excitation scheme, are able to reduce residual Doppler shifts down to $\SIangfreq{190}{\kilo \hertz}$, limited by transit time broadening~\cite{Bohaichuk2023threephoton}. Theoretically, it is possible to achieve Doppler cancellation on the order of the Rydberg state lifetimes. Therefore, these approaches are quite promising for substantially increasing sensitivity~\cite{Kuebler2018, Bohaichuk2023threephoton}. 

\begin{table}[]
\centering
\caption{Summary of the determined $\Omega_\mathrm{RF,min}$, $E_\mathrm{RF,min}$ and $S$ for the considered $\Omega_\mathrm{C}$. The bandwidth is $B= \SI{500}{\kilo \hertz}$. We use the directionally averaged dipole moment for our calculations, $\mu_\mathrm{RF} = \SI{2569.7}{\elementarycharge \bohrradius}$.}
\begin{tabular}{l|c|c|c c c|c|c}
    & \multicolumn{3}{c}{$T=0$} & \multirow{2}{10pt}{}  & \multicolumn{3}{c}{$k_\mathrm{eff}=0$}   \\ \cline{1-4} \cline{6-8}
    $\Omega_\mathrm{C}$ ($2\pi \narrowtimes \si{\mega \hertz}$)& 10 & 15 & 20 &  &10 & 15 & 20 \\ \cline{1-4} \cline{6-8} 
    $\Omega_\mathrm{RF,min}$ ($2\pi \narrowtimes \si{\kilo \hertz}$)& 1.5 & 2.2 & 2.9 &  & 17 & 13 & 12  \\ \cline{1-4} \cline{6-8}
    $E_\mathrm{RF,min}$ (\si{\micro \volt \per \centi \meter}) 
    & 0.45 & 0.69 & 0.88 &  & 5.2 & 4.0 & 3.6  \\ 
    \cline{1-4} \cline{6-8}
    $S$ (\si{\nano \volt \per \centi \meter \per \hertz\tothe{1/2}})
    & 0.63 & 0.94 & 1.2 & & 7.5 & 5.6 & 5.2  \\
    \multicolumn{7}{c}{  } \\
    & \multicolumn{3}{c}{finite $T$, finite $k_\mathrm{eff}$} & & \multicolumn{3}{c}{MOT}\\  \cline{1-4} \cline{6-8}
    $\Omega_\mathrm{C}$ ($2\pi \narrowtimes \si{\mega \hertz}$) & 10 &  15 & 20 & &10 &  15 & 20 \\ \cline{1-4} \cline{6-8}
    $\Omega_\mathrm{RF,min}$ ($2\pi \narrowtimes \si{\kilo \hertz}$) & 130 &  170 & 220 & & 13 &  19 & 25 \\ \cline{1-4} \cline{6-8}
    $E_\mathrm{RF,min}$ (\si{\micro \volt \per \centi \meter}) 
    & 40 & 52 & 63 &  & 4.0 & 5.8 & 7.5  \\ 
    \cline{1-4} \cline{6-8}
    $S$ (\si{\nano \volt \per \centi \meter \per \hertz\tothe{1/2}}) 
    & 56 & 75 & 94 & & 5.6 & 8.2 & 11
\end{tabular}
\label{Tab:Sensitivities}
\end{table}

\section{Conclusion}
In this paper, we found analytic expressions for the absorption coefficient in the AR and NL, assuming the weak probe approximation for Rydberg atom-based RF E-field sensing. We developed a picture where the optical response of the atoms can be interpreted as a modification of probe transition photon scattering. The modification depends on the coherences of each of the transitions and their respective detunings. The information about the RF E-field is transferred to the photon scattering on the probe transition via the dressing of the atom performed by the coupling laser.
With the analytic expression for the absorption coefficient, the coupling Rabi frequency, and the atomic decay rates, the incident RF E-field can be recalculated. The picture presented can be extended to other excitation schemes. On a more basic level, the results of the paper provide insight into multi-level interference and coherence phenomena in atomic and molecular spectroscopy.

We compared several different situations to elucidate how Doppler shifts affect Rydberg atom-based RF sensors. $T=0$, where all atoms are at rest, $k_\mathrm{eff}=0$, where the laser wavelengths on the probe and coupling transitions are matched, and finite temperature and wavelength mismatch, were investigated. A set of typical experimental parameters was used to compare the sensitivity to the RF field in each case for a detection bandwidth of 
$\SI{500}{\kilo\hertz}$. The finite Doppler broadening and finite wave vector mismatch describe the most common Rydberg atom-based RF sensing experiments to date. For finite wave vector mismatch in a thermal atomic gas, Doppler broadening extends the AR and NL towards larger $\Omega_\mathrm{RF}$. An analytic expression at small $\Omega_\mathrm{RF}$ was found in the strong coupling limit. In the strong coupling limit, the features of the velocity distribution can be separated and a closed form expression was found for the absorption coefficient. The expression describes how Doppler broadening decreases sensitivity. Insight was gained by comparing this expression to those obtained by assuming $T=0$  and $k_\mathrm{eff}=0$.

Sensitivities were calculated for the different cases of temperature and wave vector mismatch described in the paper. $T=0$ shows the best sensitivity to the incident RF E-field. The shot-noise limited sensitivity is estimated to be 
\SI{630}{\pico \volt \per \centi \meter \per \hertz\tothe{1/2}} at $\Omega_\mathrm{C} = \SIangfreq{10}{\mega \hertz}$. For $k_\mathrm{eff}=0$, the optical response is similar to $T=0$. However, the absorption is reduced by the mean Doppler broadening $k_\mathrm{P}\bar{v}/\gamma_2$. The shot-noise limited sensitivity is
\SI{7.5}{\nano \volt \per \centi \meter \per \hertz\tothe{1/2}} for a similar set of parameters, including atom numbers. The shot-noise limited sensitivity for finite wave vector mismatch in a thermal atomic gas was estimated to be
\SI{56}{\nano \volt \per \centi \meter \per \hertz\tothe{1/2}}. The sensitivity for finite wave vector mismatch in a thermal gas is around two orders of magnitude worse than for $T=0$, due to the additional Doppler broadening. The $k_\mathrm{eff}=0$ sensitivity is the highest when achievable experimental parameters, namely atom number, are considered, since many more atoms can be accessed in a vapor cell than a MOT. The fact that narrow spectral line widths are advantageous for achieving optimum sensitivity in thermal vapor cells in Rydberg atom-based sensing implies that the three photon approach to Rydberg atom-based RF sensing described in \cite{Kuebler2018, Bohaichuk2023threephoton} is a promising, practical path forward to significantly improve sensitivity.

\section{Acknowledgements}
The authors thank Defence Research and Development Canada through the Ideas program (W7714-228077) and Defense Advanced Research Projects Agency through the Savant program (HR001120S0062) for support.




\bibliography{biblio_v2}

\end{document}